\documentclass[twocolumn]{aastex701}
\let\tablenum\relax

\usepackage{url}
\usepackage{threeparttable}
\usepackage{ulem}
\usepackage{natbib}
\usepackage{appendix}
\usepackage{amsmath}
\usepackage{amssymb}
\usepackage{multirow}
\usepackage{float}
\usepackage{physics}
\usepackage{siunitx}
\usepackage{mhchem}
\usepackage{appendix}

\def\ion#1#2{#1$\;${\sc\@roman{#2}}\relax}
\def\lesssim{\mathrel{\hbox{\rlap{\hbox{\lower4pt\hbox{$\sim$}}}\hbox{$<$}}}}
\def\gtrsim{\mathrel{\hbox{\rlap{\hbox{\lower4pt\hbox{$\sim$}}}\hbox{$>$}}}}

\def\red{\textcolor{black}}

\newcommand{\arcsecond}{\ensuremath{''}}

\shorttitle{Strongly Lensed Clumpy Galaxy at $z\sim11-12$}
\shortauthors{Nakane et al.}
\begin{document}
\title{
VENUS: A Strongly Lensed Clumpy Galaxy at $z\sim11-12$\\
behind the Galaxy Cluster MACS J0257.1-2325
%VENUS: A Strongly Lensed Galaxy at $z\sim11-12$\\
%behind the Galaxy Cluster MACS J0257.1-2325
%VENUS: A Strongly Lensed $z\sim12$ Galaxy \\
%Resolved into Multiple Clumps
}

\author[0009-0000-1999-5472]{Minami Nakane}
\affiliation{Institute for Cosmic Ray Research, The University of Tokyo, 5-1-5 Kashiwanoha, Kashiwa, Chiba 277-8582, Japan}
\affiliation{Department of Physics, Graduate School of Science, The University of Tokyo, 7-3-1 Hongo, Bunkyo, Tokyo 113-0033, Japan}
\email[show]{nakanem@icrr.u-tokyo.ac.jp}

\author[0000-0002-5588-9156]{Vasily Kokorev}
\affiliation{Department of Astronomy, The University of Texas at Austin, 2515 Speedway, Stop C1400, Austin, TX 78712, USA} 
\affiliation{Cosmic Frontier Center, The University of Texas at Austin, Austin, TX 78712, USA}
\email{vasily.kokorev.astro@gmail.com}

\author[0000-0001-7201-5066]{Seiji Fujimoto}
\affiliation{David A. Dunlap Department of Astronomy and Astrophysics, University of Toronto, 50 St. George Street, Toronto, Ontario, M5S 3H4, Canada}
\affiliation{Dunlap Institute for Astronomy and Astrophysics, 50 St. George Street, Toronto, Ontario, M5S 3H4, Canada}
\email{seiji.fujimoto@utoronto.ca}

\author[0000-0002-1049-6658]{Masami Ouchi}
\affiliation{National Astronomical Observatory of Japan, 2-21-1 Osawa, Mitaka, Tokyo 181-8588, Japan}
\affiliation{Institute for Cosmic Ray Research, The University of Tokyo, 5-1-5 Kashiwanoha, Kashiwa, Chiba 277-8582, Japan}
\affiliation{Department of Astronomical Science, SOKENDAI (The Graduate University for Advanced Studies), 2-21-1 Osawa, Mitaka, Tokyo, 181-8588, Japan}
\affiliation{Kavli Institute for the Physics and Mathematics of the Universe (WPI), The University of Tokyo, 5-1-5 Kashiwanoha, Kashiwa, Chiba 277-8583, Japan}
\email{ouchims@icrr.u-tokyo.ac.jp}

\author[0000-0003-4368-3326]{Derek J. McLeod}
\affiliation{Institute for Astronomy, University of Edinburgh, Royal Observatory, Edinburgh EH9 3HJ, UK}
\email{derek.mcleod@ed.ac.uk}

\author[0000-0001-9411-3484]{Miriam Golubchik}
\affiliation{Department of Physics, Ben-Gurion University of the Negev, P.O. Box 653, Beer-Sheva 8410501, Israel}
\email{golubmir@post.bgu.ac.il}

\author[0000-0003-3484-399X]{Masamune Oguri}
\affiliation{Center for Frontier Science, Chiba University, 1-33 Yayoi-cho, Inage-ku, Chiba 263-8522, Japan}
\affiliation{Department of Physics, Graduate School of Science, Chiba University, 1-33 Yayoi-Cho, Inage-Ku, Chiba 263-8522, Japan}
\email{masamune.oguri@chiba-u.jp}

\author[0000-0002-0350-4488]{Adi Zitrin}
\affiliation{Department of Physics, Ben-Gurion University of the Negev, P.O. Box 653, Beer-Sheva 8410501, Israel}
\email{adizitrin@gmail.com}

\author[0009-0008-3775-5112]{Cecilia Bondestam}
\affiliation{Institute for Astronomy, University of Edinburgh, Royal Observatory, Edinburgh EH9 3HJ, UK}
\email{C.Bondestam@ed.ac.uk}

\author[0000-0002-7622-0208]{Callum T. Donnan}
\affiliation{NSF's National Optical-Infrared Astronomy Research Laboratory, 950 N. Cherry Ave., Tucson, AZ 85719, USA}
\email{callum.donnan@noirlab.edu}

\author[0000-0003-2680-005X]{Gabriel Brammer}
\affiliation{Cosmic Dawn Center (DAWN), Denmark}
\affiliation{Niels Bohr Institute, University of Copenhagen, Jagtvej 128, 2200 Copenhagen N, Denmark}
\email{gabriel.brammer@nbi.ku.dk}

\author[0000-0001-8519-1130]{Steven L. Finkelstein}
\affiliation{Department of Astronomy, The University of Texas at Austin, 2515 Speedway, Stop C1400, Austin, TX 78712, USA} 
\affiliation{Cosmic Frontier Center, The University of Texas at Austin, Austin, TX 78712, USA}
\email{stevenf@astro.as.utexas.edu}

\author[0000-0002-4201-7367]{Chris Willott} \affiliation{NRC Herzberg, 5071 West Saanich Rd, Victoria, BC V9E 2E7, Canada}
\email{chris.willott@nrc.ca}

\author[0009-0009-4388-898X]{Gregor Rihtarsic}
\affiliation{University of Ljubljana, Faculty of Mathematics and Physics, Jadranska ulica 19, SI-1000 Ljubljana, Slovenia}
\email{gregor.rihtarsic@fmf.uni-lj.si}

\author[0000-0001-8325-1742]{Guillaume Desprez}
\affiliation{Kapteyn Astronomical Institute, University of Groningen, P.O. Box 800, 9700AV Groningen, The Netherlands}
\email{g.p.a.desprez@rug.nl}

\author[0000-0002-8192-8091]{Angela Adamo}
\affiliation{Department of Astronomy, The Oskar Klein Centre, Stockholm University, AlbaNova, SE-10691 Stockholm, Sweden}
\email{angela.adamo@astro.su.se}

\author[0000-0002-5057-135X]{Eros Vanzella}
\affiliation{INAF -- OAS, Osservatorio di Astrofisica e Scienza dello Spazio di Bologna, via Gobetti 93/3, I-40129 Bologna, Italy}
\email{eros.vanzella@inaf.it}

\author[0000-0001-5984-0395]{Maru\v{s}a Brada\v{c}}
\affiliation{University of Ljubljana, Department of Mathematics and Physics, Jadranska ulica 19, SI-1000 Ljubljana, Sloveni}
\affiliation{Department of Physics and Astronomy, University of California Davis, 1 Shields Avenue, Davis, CA 95616, USA}
\email{marusa.bradac@fmf.uni-lj.si}

\author[0000-0003-1427-2456]{Matteo Messa}
\affiliation{INAF -- OAS, Osservatorio di Astrofisica e Scienza dello Spazio di Bologna, via Gobetti 93/3, I-40129 Bologna, Italy}
\email{matteo.messa@inaf.it}

\author[0009-0006-6763-4245]{Hiroto Yanagisawa}
\affiliation{Institute for Cosmic Ray Research, The University of Tokyo, 5-1-5 Kashiwanoha, Kashiwa, Chiba 277-8582, Japan}
\affiliation{Department of Physics, Graduate School of Science, The University of Tokyo, 7-3-1 Hongo, Bunkyo, Tokyo 113-0033, Japan}
\email{yana@icrr.u-tokyo.ac.jp}

\author[0000-0002-4622-6617]{Fengwu Sun}
\affiliation{Center for Astrophysics $|$ Harvard \& Smithsonian, 60 Garden St., Cambridge, MA 02138, USA}
\email{fengwu.sun@cfa.harvard.edu}

\author[0000-0001-7113-2738]{Henry C. Ferguson} \affiliation{Space Telescope Science Institute, 3700 San Martin Drive, Baltimore, MD 21218, USA}
\email{ferguson@stsci.edu}

\author[0000-0003-1581-7825]{Ray A. Lucas} \affiliation{Space Telescope Science Institute, 3700 San Martin Drive, Baltimore, MD 21218, USA}
\email{lucas@stsci.edu}

\newcommand{\STScI}{\affiliation{Space Telescope Science Institute (STScI), 3700 San Martin Drive, Baltimore, MD 21218, USA}}
\newcommand{\JHU}{\affiliation{Center for Astrophysical Sciences, Department of Physics and Astronomy, The Johns Hopkins University, 3400 N Charles St. Baltimore, MD 21218, USA}}
\newcommand{\ESAAURA}{\affiliation{Association of Universities for Research in Astronomy (AURA), Inc.~for the European Space Agency (ESA)}}
\author[0000-0001-7410-7669]{Dan Coe} 
\STScI 
\ESAAURA 
\JHU
\email{dcoe@stsci.edu}

\author[0000-0001-5492-1049]{Johan Richard}
\affiliation{Univ Lyon, Univ Lyon 1, Ens de Lyon, CNRS, Centre de Recherche Astrophysique de Lyon UMR5574, F-69230, Saint-Genis-Laval, France}
\email{johan.richard@univ-lyon1.fr}

\author[0000-0002-5258-8761]{Abdurro'uf} 
\affiliation{Department of Astronomy, Indiana University, 727 East Third Street, Bloomington, IN 47405, USA}
\email{fnuabdur@iu.edu}

\author[0000-0003-3596-8794]{Hollis B. Akins}
\affiliation{Department of Astronomy, The University of Texas at Austin, 2515 Speedway, Stop C1400, Austin, TX 78712, USA} 
\email{hollis.akins@gmail.com}

\author[0000-0003-2718-8640]{Joseph F. V. Allingham}
\affiliation{Department of Physics, Ben-Gurion University of the Negev, P.O. Box 653, Be'er-Sheva 84105, Israel}
\email{allingha@post.bgu.ac.il}

\author[0000-0001-5758-1000]{Ricardo O. Amor\'{i}n} 
\affiliation{Instituto de Astrof\'isica de Andaluc\'ia--CSIC, Glorieta de la Astronom\'ia s/n, E--18008 Granada, Spain}
\email{amorin@iaa.es}

\author[0000-0003-3983-5438]{Yoshihisa Asada}
\affiliation{Dunlap Institute for Astronomy and Astrophysics, 50 St. George Street, Toronto, Ontario, M5S 3H4, Canada}
\email{yoshi.asada@utoronto.ca}

\author[0000-0002-7570-0824]{Hakim Atek}
\affiliation{Institut d'Astrophysique de Paris, CNRS, Sorbonne Universit\'e, 98bis Boulevard Arago, 75014, Paris, France}
\email{atek@iap.fr}

\author[0000-0002-8686-8737]{Franz E. Bauer}
\affiliation{Instituto de Alta Investigaci{\'{o}}n, Universidad de Tarapac{\'{a}}, Casilla 7D, Arica, 1010000, Chile}
\email{franz.e.bauer@gmail.com}

\author[0000-0001-5063-8254]{Rachel Bezanson}
\affiliation{Department of Physics and Astronomy and PITT PACC, University of Pittsburgh, Pittsburgh, PA 15260, USA}
\email{rachel.bezanson@pitt.edu}

\author[0000-0002-7908-9284]{Larry D. Bradley}
\affiliation{Space Telescope Science Institute, 3700 San Martin Drive, Baltimore, MD 21218, USA}
\email{lbradley@stsci.edu}

\author[0000-0002-0302-2577]{John Chisholm}
\affiliation{Department of Astronomy, The University of Texas at Austin, 2515 Speedway, Stop C1400, Austin, TX 78712, USA} 
\affiliation{Cosmic Frontier Center, The University of Texas at Austin, Austin, TX 78712, USA}
\email{chisholm@austin.utexas.edu}

\author[0000-0003-1949-7638]{Christopher J. Conselice}
\affiliation{Jodrell Bank Centre for Astrophysics, University of Manchester, Oxford Road, Manchester M13 9PL, UK}
\email{conselice@gmail.com}

\author[0000-0001-8460-1564]{Pratika Dayal}     
\affiliation{Canadian Institute for Theoretical Astrophysics, 60 St George St, University of Toronto, Toronto, ON M5S 3H8, Canada}
\affiliation{David A. Dunlap Department of Astronomy and Astrophysics, University of Toronto, 50 St George St, Toronto ON M5S 3H4, Canada}
\affiliation{Department of Physics, 60 St George St, University of Toronto, Toronto, ON M5S 3H8, Canada}
\email{pdayal@cita.utoronto.ca}

\author[0000-0003-0348-2917]{Miroslava Dessauges-Zavadsky} 
\affiliation{Department of Astronomy, University of Geneva, Chemin Pegasi 51, 1290 Versoix, Switzerland}
\email{miroslava.dessauges@unige.ch}

\author[0000-0001-9065-3926]{Jose M. Diego}
\affiliation{Instituto de F\'{i}sica de Cantabria (CSIC-UC), Avenida. Lastros s/n. 39005 Santander, Spain}
\email{jdiego@ifca.unican.es}

\author[0000-0002-9382-9832]{Andreas L. Faisst}
\affiliation{IPAC, California Institute of Technology, 1200 E. California Blvd. Pasadena, CA 91125, USA}
\email{afaisst@caltech.edu}

\author[0009-0007-2578-9238]{Gavin Farley}
\affiliation{David A. Dunlap Department of Astronomy \& Astrophysics, University of Toronto, 50 St. George St, Toronto, ON M5S 3H4, Canada}
\email{gavin.farley@mail.utoronto.ca}

\author[0000-0001-7232-5355]{Qinyue Fei}
\affiliation{David A. Dunlap Department of Astronomy and Astrophysics, University of Toronto, 50 St. George Street, Toronto, Ontario, M5S 3H4, Canada}
\email{qyfei.astro@gmail.com}

\author[0000-0003-1625-8009]{Brenda L. Frye}
\affiliation{Department of Astronomy/Steward Observatory, University of Arizona, 933 N. Cherry Avenue, Tucson, AZ 85721, USA}
\email{brendafrye@gmail.com}

\author[0000-0001-7440-8832]{Yoshinobu Fudamoto}
\affiliation{Center for Frontier Science, Chiba University, 1-33 Yayoi-cho, Inage-ku, Chiba 263-8522, Japan}
\email{yoshinobu.fudamoto@gmail.com}

\author[0000-0001-6278-032X]{Lukas J. Furtak}
\affiliation{Cosmic Frontier Center, The University of Texas at Austin, Austin, TX 78712, USA} 
\affiliation{Department of Astronomy, The University of Texas at Austin, Austin, TX 78712, USA} 
\email{furtak@utexas.edu}

\author[0000-0002-6047-430X]{Yuichi Harikane} 
\affiliation{Institute for Cosmic Ray Research, The University of Tokyo, 5-1-5 Kashiwanoha, Kashiwa, Chiba 277-8582, Japan}
\email{hari@icrr.u-tokyo.ac.jp}

\author[0000-0003-4512-8705]{Tiger Yu-Yang Hsiao}  
\affiliation{Department of Astronomy, The University of Texas at Austin, 2515 Speedway, Stop C1400, Austin, TX 78712, USA} 
\affiliation{Cosmic Frontier Center, The University of Texas at Austin, Austin, TX 78712, USA}
\email{tiger.hsiao@utexas.edu}

\author[0000-0002-6090-2853]{Yolanda Jim\'enez-Teja}
\affiliation{Instituto de Astrof\'isica de Andaluc\'ia--CSIC, Glorieta de la Astronom\'ia s/n, E--18008 Granada, Spain}
\affiliation{Observat\'orio Nacional, Rua General Jos\'e Cristino, 77 - Bairro Imperial de S\~ao Crist\'ov\~ao, Rio de Janeiro, 20921-400, Brazil}
\email{yojite@iaa.es}

\author[0000-0001-9187-3605]{Jeyhan S. Kartaltepe}
\affiliation{Laboratory for Multiwavelength Astrophysics, School of Physics and Astronomy, Rochester Institute of Technology, 84 Lomb Memorial Drive, Rochester, NY 14623, USA}
\email{jeyhan@astro.rit.edu}

\author[0009-0004-4332-9225]{Tomokazu Kiyota}
\affiliation{Department of Astronomical Science, The Graduate University for Advanced Studies, SOKENDAI, 2-21-1 Osawa, Mitaka, Tokyo, 181-8588, Japan}
\affiliation{National Astronomical Observatory of Japan, 2-21-1 Osawa, Mitaka, Tokyo, 181-8588, Japan}
\email{tomokazu.kiyota@grad.nao.ac.jp}

\author[0000-0002-6610-2048]{Anton M. Koekemoer}
\affiliation{Space Telescope Science Institute, 3700 San Martin Drive, Baltimore, MD 21218, USA}
\email{koekemoer@stsci.edu}

\author[0000-0003-3021-8564]{Claudia del P. Lagos}
\affiliation{International Centre for Radio Astronomy Research, The University of Western Australia, 35 Stirling Highway, Crawley, WA6009, Australia}
\affiliation{ARCCentre of Excellence for All Sky Astrophysics in 3 Dimensions (ASTRO 3D), Canberra, ACT 2611, Australia}
\email{claudia.lagos@uwa.edu.au}

\author[0000-0002-4872-2294]{Georgios E. Magdis}
\affiliation{Cosmic Dawn Center (DAWN), Denmark}
\affiliation{DTU-Space, Technical University of Denmark, Elektrovej 327, 2800, Kgs. Lyngby, Denmark}
\email{geoma@space.dtu.dk}

\author[0000-0002-7876-4321]{Ashish Kumar Meena}
\affiliation{Department of Physics, Indian Institute of Science, Bengaluru 560012, India}
\email{ashishmeena766@gmail.com}

\author[0000-0002-8530-9765]{Lamiya Mowla}
\affiliation{Department of Physics and Astronomy, Wellesley College, Wellesley, MA 02481, USA}
\email{lbamowla@gmail.com}

\author{Ga\"el Noirot} 
\affiliation{Space Telescope Science Institute, 3700 San Martin Drive, Baltimore, MD 21218, USA}
\email{gael.noirot@gmail.com}

\author[0000-0001-5851-6649]{Pascal A. Oesch}
\affiliation{Department of Astronomy, University of Geneva, Chemin Pegasi 51, 1290 Versoix, Switzerland}
\affiliation{Cosmic Dawn Center (DAWN), Denmark}
\email{pascal.oesch@unige.ch}

\author[0000-0001-9011-7605]{Yoshiaki Ono}
\affiliation{Institute for Cosmic Ray Research, The University of Tokyo, 5-1-5 Kashiwanoha, Kashiwa, Chiba 277-8582, Japan}
\email{ono@icrr.u-tokyo.ac.jp}

\author[0000-0002-6150-833X]{Rafael Ortiz III}
\affiliation{School of Earth and Space Exploration, Arizona State University, Tempe, AZ 85287-6004, USA}
\email{rortizii@asu.edu}

\author[0000-0002-9651-5716]{Richard Pan}
\affiliation{Department of Physics \& Astronomy, Tufts University, Medford, MA 02155, USA}
\email{richard.pan@tufts.edu}

\author[0000-0001-7503-8482]{Casey Papovich}
\affiliation{Department of Physics and Astronomy, Texas A\&M University, College Station, TX, 77843-4242 USA}
\affiliation{George P. and Cynthia Woods Mitchell Institute for Fundamental Physics and Astronomy, Texas A\&M University, College Station, TX, 77843-4242 USA}
\email{papovich@tamu.edu}

\author[0000-0002-2361-7201]{Justin D. R. Pierel}
\affiliation{Space Telescope Science Institute, 3700 San Martin Drive, Baltimore, MD 21218, USA}
\email{jpierel@stsci.edu}

\author[0000-0003-4223-7324]{Massimo Ricotti}
\affiliation{Department of Astronomy, University of Maryland, College Park, 20742, USA}
\email{ricotti@astro.umd.edu}

\author[0000-0002-6265-2675]{Luke Robbins}
\affiliation{Department of Physics and Astronomy, Tufts University, Medford, MA 02155, USA}
\email{lukerobbins3@gmail.com}

\author[0000-0001-7144-7182]{Daniel Schaerer}
\affiliation{Department of Astronomy, University of Geneva, Chemin Pegasi 51, 1290 Versoix, Switzerland}
\affiliation{CNRS, IRAP, 14 Avenue Edouard Belin, 31400 Toulouse, France}
\email{daniel.schaerer@unige.ch}

\author[0000-0001-9317-2888]{Raffaella Schneider}
\affiliation{Department of Physics, Sapienza University of Rome, Pzz.le Aldo Moro 5, 00185 Rome, Italy}
\email{raffaella.schneider@uniroma1.it}

\author[0000-0002-8460-0390]{Tommaso Treu}
\affiliation{Department of Physics and Astronomy, University of California, Los Angeles, CA, 90095, USA}
\email{tt@astro.ucla.edu}

\author[0000-0001-6477-4011]{Francesco Valentino}
\affiliation{Cosmic Dawn Center (DAWN), Denmark}
\affiliation{DTU Space, Technical University of Denmark, Elektrovej 327, DK-2800 Kgs. Lyngby, Denmark}
\email{fmava@dtu.dk}

\author[0000-0001-8156-6281]{Rogier A. Windhorst}
\affiliation{School of Earth and Space Exploration, Arizona State University, Tempe, AZ 85287-6004, USA}
\email{Rogier.Windhorst@asu.edu}

\author[0000-0003-0212-2979]{Volker Bromm}
\affiliation{Department of Astronomy, The University of Texas at Austin, 2515 Speedway, Stop C1400, Austin, TX 78712, USA} 
\affiliation{Cosmic Frontier Center, The University of Texas at Austin, Austin, TX 78712, USA}
\affiliation{Weinberg Institute for Theoretical Physics, Texas Center for Cosmology and Astroparticle Physics, University of Texas at Austin, Austin, TX 78712, USA}
\email{vbromm@astro.as.utexas.edu}

\author[0000-0003-1344-9475]{Eiichi Egami}
\affiliation{Steward Observatory, University of Arizona, 933 N. Cherry Avenue, Tucson, AZ85721, USA}
\email{egami@arizona.edu}

\author[0000-0002-4837-1615]{Mauro Gonz\'alez-Otero}
\affiliation{Instituto de Astrof\'isica de Andaluc\'ia--CSIC, Glorieta de la Astronom\'ia s/n, E--18008 Granada, Spain}
\email{mauromarago@gmail.com}

\author[0000-0002-4052-2394]{Kotaro Kohno}
\affiliation{Institute of Astronomy, Graduate School of Science, The University of Tokyo, 2-21-1 Osawa, Mitaka, Tokyo, 181-0015 Japan}
\affiliation{Research Center for the Early Universe, Graduate School of Science, The University of Tokyo, 7-3-1 Hongo, Bunkyo-ku, Tokyo 113-0033, Japan}
\email{kkohno@ioa.s.u-tokyo.ac.jp}

\author[0000-0002-2057-5376]{Ivo Labbe}
\affiliation{Centre for Astrophysics and Supercomputing, Swinburne University of Technology, Melbourne, VIC 3122, Australia}
\email{ivolabbe@gmail.com}

\author[0000-0003-2871-127X]{Jorryt Matthee}  \affiliation{Institute of Science and Technology Austria (ISTA), Am Campus 1, 3400 Klosterneuburg, Austria}
\email{jorryt.matthee@ist.ac.at}

\author[0000-0002-3706-9955]{Marcie Mun}
\affiliation{Institut d'Astrophysique de Paris, CNRS, Sorbonne Universit\'e, 98bis Boulevard Arago, 75014, Paris, France}
\email{mun@iap.fr}

\author[0000-0003-3997-5705]{Rohan P. Naidu}
\affiliation{MIT Kavli Institute for Astrophysics and Space Research, 70 Vassar Street, Cambridge, MA 02139, USA}
\email{rnaidu@mit.edu}

\author[0000-0002-9909-3491]{Roberta Tripodi}
\affiliation{INAF -- Osservatorio Astronomico di Roma, Via Frascati 33, I-00078 Monte Porzio Catone, Italy}
\affiliation{University of Ljubljana FMF, Jadranska 19, 1000 Ljubljana, Slovenia}
\affiliation{IFPU, Institute for Fundamental Physics of the Universe, Via Beirut 2, 34014 Trieste, Italy}
\email{roberta.tripodi@fmf.uni-lj.si}

%% Note that the \and command from previous versions of AASTeX is now
%% depreciated in this version as it is no longer necessary. AASTeX 
%% automatically takes care of all commas and "and"s between authors names.

%% AASTeX 6.31 has the new \collaboration and \nocollaboration commands to
%% provide the collaboration status of a group of authors. These commands 
%% can be used either before or after the list of corresponding authors. The
%% argument for \collaboration is the collaboration identifier. Authors are
%% encouraged to surround collaboration identifiers with ()s. The 
%% \nocollaboration command takes no argument and exists to indicate that
%% the nearby authors are not part of surrounding collaborations.

%% Mark off the abstract in the ``abstract'' environment. 
\begin{abstract}
We present the discovery of a strongly lensed galaxy at $z\sim11$--$12$, dubbed the ``Misty Moons'', identified in the JWST Treasury Survey, Vast Exploration for Nascent, Unexplored Sources (VENUS). The Misty Moons is gravitationally lensed by the galaxy cluster MACS J0257.1-2325 at $z=0.505$, and has five multiple images suggested by two independent lens models. Two of the five images, ID1 and ID2 ($\mu\sim 20-30$), are very bright (F200W$\sim26$ AB mag) and exhibit blue SEDs with prominent Ly$\alpha$ breaks. In the source plane, the Misty Moons is a sub-$L^*$ galaxy ($M_{\rm UV}\sim-18.0$ mag) resolved into multiple stellar clumps, each of which has an effective radius of $r_\mathrm{eff}\sim 10$--$70$ pc and a stellar mass of $\sim10^7\ M_\odot$. These clumps dominate the stellar mass budget of the Misty Moons ($\gtrsim80\%$), similar to other high-$z$ clumps, which suggests a highly clustered mode of star formation in the early Universe, unlike seen in local dwarf galaxies. We convolve the source-plane image with the JWST/NIRCam point-spread function to produce a mock NIRCam image of the Misty Moons without lensing magnification, and find that the intrinsic galaxy has a radial surface-brightness profile comparable to those of $z\gtrsim10$ faint galaxies, such as JADES-GS-z13-0 and JADES-GS-z14-1, indicating that the Misty Moons represents a typical $z\gtrsim10$ faint galaxy. The Misty Moons, a lensed galaxy with resolved internal structures, provides an ideal laboratory for exploring the early stages of galaxy formation at $z\gtrsim10$.

\end{abstract}

%% Keywords should appear after the \end{abstract} command. 
%% The AAS Journals now uses Unified Astronomy Thesaurus concepts:
%% https://astrothesaurus.org
%% You will be asked to selected these concepts during the submission process
%% but this old "keyword" functionality is maintained in case authors want
%% to include these concepts in their preprints.

%% From the front matter, we move on to the body of the paper.
%% Sections are demarcated by \section and \subsection, respectively.
%% Observe the use of the LaTeX \label
%% command after the \subsection to give a symbolic KEY to the
%% subsec tion for cross-referencing in a \ref command.
%% You can use LaTeX's \ref and \label commands to keep track of
%% cross-references to sections, equations, tables, and figures.
%% That way, if you change the order of any elements, LaTeX will
%% automatically renumber them.
%%
%% We recommend that authors also use the natbib \citep
%% and \citet commands to identify citations.  The citations are
%% tied to the reference list via symbolic KEYs. The KEY corresponds
%% to the KEY in the \bibitem in the reference list below. 

\section{Introduction}
\label{sec:introduction}

JWST observations have been unveiling the nature of early galaxies up to $z>10$ (e.g., \citealt{ArrabalHaro2023,Curtis-Lake2023,Carniani2024,Naidu2025}). Some of the intrinsically bright galaxies at $z\gtrsim6$, such as GHZ2 ($z_\mathrm{spec}=12.34$; \citealt{Castellano2024,Zavala2024}) and GN-z11 ($z_\mathrm{spec}=10.60$; \citealt{Bunker2023,Maiolino2024}), show compact morphologies with half-light radii of $r_\mathrm{eff}\lesssim100$ pc and nitrogen-rich abundance ratios, suggesting a possible connection with globular clusters (GCs; e.g., \citealt{Isobe2023,Isobe2025,Senchyna2024,Topping2024,Topping2025,Harikane2025,Ji2025}). However, the stellar masses of these galaxies are more than one order of magnitude larger than typical local GCs, which suggests possibilities of superposition of several individual clusters and significant mass loss due to dynamical evolution (e.g., \citealt{Webb2015,Zavala2024}). In addition, these galaxies are not resolved into their components, which complicates a direct comparison. Gravitational lensing bridges this gap by boosting both signal-to-noise (S/N) ratio and spatial resolution. High magnification by gravitational lensing has enabled identifications of $1-10$ pc scale star clusters or stellar clumps with stellar masses of $M_*\sim10^4-10^7M_\odot$ at $z\gtrsim6$ (e.g., \citealt{Vanzella2023,Adamo2024,Bradley2024,Mowla2024,Fujimoto2025,Abdurrouf2025,Bradac2025,Messa2025,Messa2026}). Overall, they share comparable properties of young stellar age, low metallicity, and low dust extinction. They also exhibit high stellar mass surface densities ($\Sigma_*$), some of which have about three orders of magnitude higher ($\Sigma_*>10^5\ M_\odot\ \mathrm{pc^{-2}}$) than typical young star clusters in the local Universe \citep{Brown2021}.

Star cluster formation in the local Universe is quantitatively evaluated by the cluster formation efficiency (CFE), which is a fraction of total stellar mass of star clusters to that of their host galaxy, formed in a given time interval \citep[e.g., ][]{Goddard2010,Adamo2011,Adamo2020,Kruijssen2012,Messa2018,Cook2023}. These studies have suggested a positive correlation between the CFE and star formation rate (SFR) surface density ($\Sigma_\mathrm{SFR}$). An increasing trend of $\Sigma_\mathrm{SFR}$ with redshift for high-$z$ star-forming galaxies \citep[e.g., ][]{Morishita2024} indicates that the CFE in the early Universe could be high. A recent study of \citet{Vanzella2026} has investigated the CFE of the Cosmic Gems galaxy at $z_\mathrm{spec}=9.63$, which contains $1$ pc scale gravitationally bound star clusters \citep{Adamo2024,Bradley2024,Messa2026}. They conclude that the individual star clusters within the Cosmic Gems galaxy require a large CFE ($\Gamma=50–100\%$), together with a top-heavy star cluster mass function or a slope $\beta>-2$ with a higher minimum cluster mass limit, in order to reproduce the observed massive clusters. Currently, the detections of high-$z$ galaxies, which are resolved into star cluster scales, remain rare, especially beyond $z\sim10$, due to the relatively small volume probed at sufficiently high magnification required to observe such intrinsically faint objects. This motivates a search for highly magnified $z>10$ galaxies across as many strong-lensing cluster fields as possible, to explore in detail the star cluster and clump formation in the early Universe.

\begin{figure*}[ht!]
    \centering
    \includegraphics[width=0.9\linewidth]{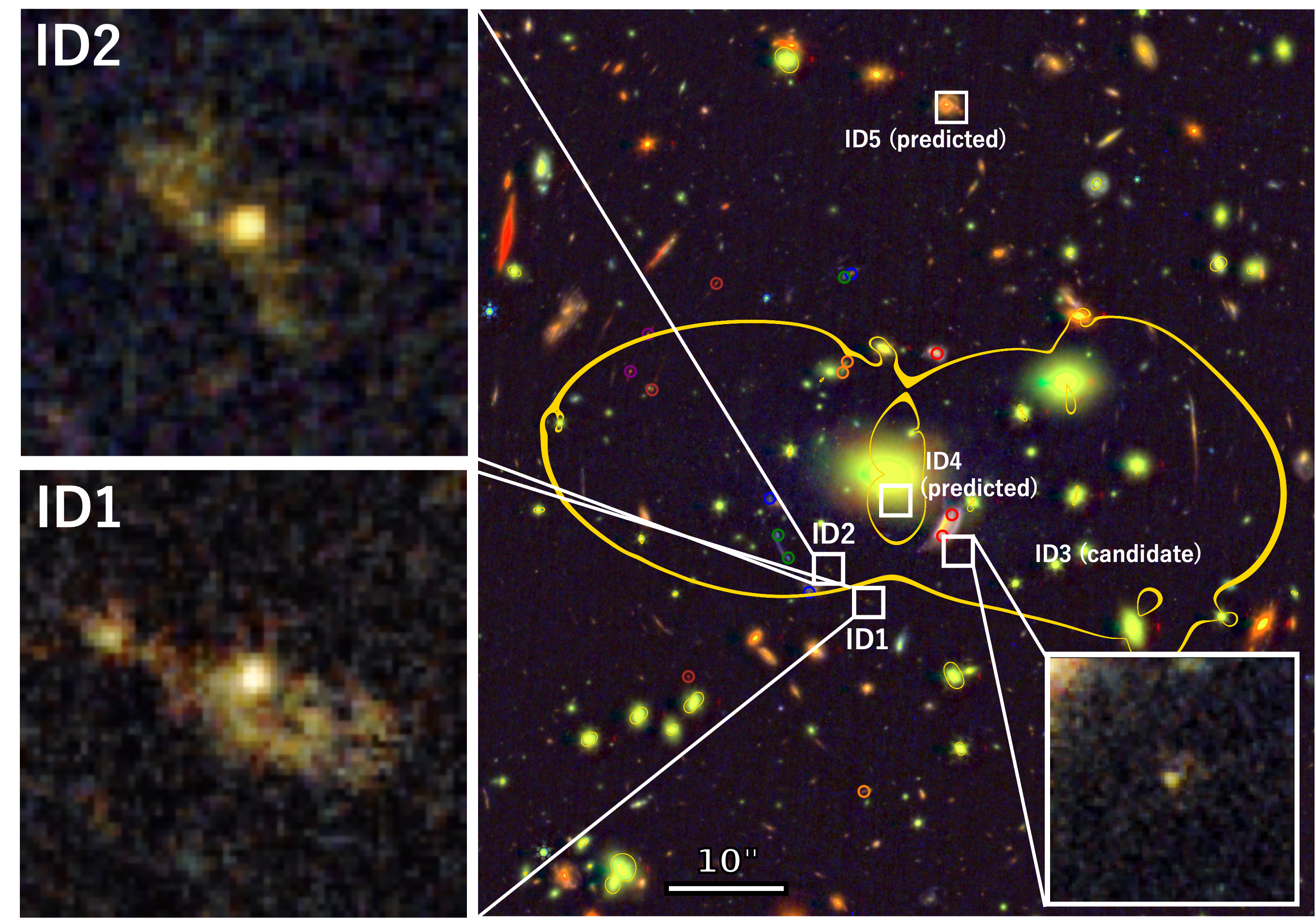}
    \caption{NIRCam RGB color image (\red{R: F356W+F410M+F444W, G: F150W+F200W+F277W, B: F435W+F814W+F150W}) of the MACS J0257.1-2325 cluster field. The yellow curve indicates the critical curve at $z=11-12$. The white squares show the positions of ID1-ID5 of the Misty Moons predicted by our independent lens models while the colored circles denote the $z\sim1-8$ multiple image systems used to constrain the lens models (see Section \ref{sec:lensmodel}). $2\arcsecond\times2\arcsecond$ cutout images show the zoom in on ID1, ID2, and a candidate for ID3.}
    \label{fig:Misty_Moons}
\end{figure*}

In this work, we present the discovery of a strongly lensed and highly magnified clumpy galaxy at $z_\mathrm{phot}\sim11-12$, dubbed the ``Misty Moons'' (Figure \ref{fig:Misty_Moons}), identified in the Vast Exploration for Nascent, Unexplored Sources (VENUS) JWST/NIRCam images and the Reionization Lensing Cluster Survey (RELICS) HST images of the lensing galaxy cluster MACS J0257.1-2325 at $z=0.505$. This paper is organized as follows. In section \ref{sec:obs_and_photometry}, we describe our observations and photometric measurements. Section \ref{sec:lensmodel} details our lens models based on the latest NIRCam observations. In Section \ref{sec:analysis}, we explain the analysis procedures including clump modeling and spectral energy distribution (SED) fitting. In Section \ref{sec:res_diss}, we present our results, and discuss the physical properties and clump formation of the Misty Moons. Section \ref{sec:summary} summarizes our findings. Throughout this paper, we assume a standard $\Lambda$CDM cosmology with $\Omega_\Lambda=0.7$, $\Omega_m=0.3$, and $H_0=70$ km $\mathrm{s}^{-1}$ $\mathrm{Mpc}^{-1}$. All magnitudes are in the AB system \citep{Oke&Gunn1983}. 

\section{Observation and Photometry} 
\label{sec:obs_and_photometry}

\setcounter{footnote}{0}
\subsection{JWST and HST Observations}
\label{subsec:obs}
The galaxy cluster MACS J0257.1-2325 (hereafter MACS0257-2325), located at $z=0.505$, was observed with JWST/NIRCam as part of the treasury lensing cluster survey VENUS (Cycle 4 GO-6882, PIs: S. Fujimoto \& D. Coe; Fujimoto et al. in preparation) on 18-19 August 2025. The NIRCam observations utilize ten bands consisting of F090W, F115W, F150W, F200W, F210M, F277W, F300M, F356W, F410M, and F444W with exposure times of $0.35-0.57$ hours to homogeneously achieve a source detection limit of $\sim$28 mag across all the NIRCam filters (5$\sigma$, point source). A dedicated paper describing the VENUS data reduction procedure is forthcoming. Here, we only briefly describe our methodology. We start with JWST level-2 products from the Mikulski Archive for
Space Telescopes (MAST) and reduce them with the \texttt{grizli} pipeline \citep{Brammer2021,Brammer2023}, similarly to the widely available public DAWN JWST Archive (DJA) products\footnote{\url{https://dawn-cph.github.io/dja/}}. The photometric calibration was performed with the Calibration Reference Data System (CRDS) context \texttt{jwst\_1456.pmap}. The \texttt{grizli} procedure implements crucial improvements over the standard STScI pipeline, including corrections for cosmic rays, stray light, and detector artifacts \citep{Bradley2023,Rigby2023}. Further, we implement additional background, 1/f noise and diffraction spike subtraction procedures, both at the amplifier level, for each filter, and then the final drizzled mosaic \citep[e.g., see][]{Endsley2023,Kokorev2025a}. JWST NIRCam images are drizzled to a 0\farcs{03}/pix grid. We also incorporate HST imaging of MACS J0257.1-2325 \citep{Coe2019} obtained with ACS/WFC F435W, F555W, and F814W. The HST images are based on Gaia-aligned mosaics from the CHArGE archive \citep{Kokorev2022}, which are then drizzled on the same footprint and pixel scale as the JWST data. 

\begin{figure*}[ht!]
    \centering
    \includegraphics[width=0.90\linewidth]{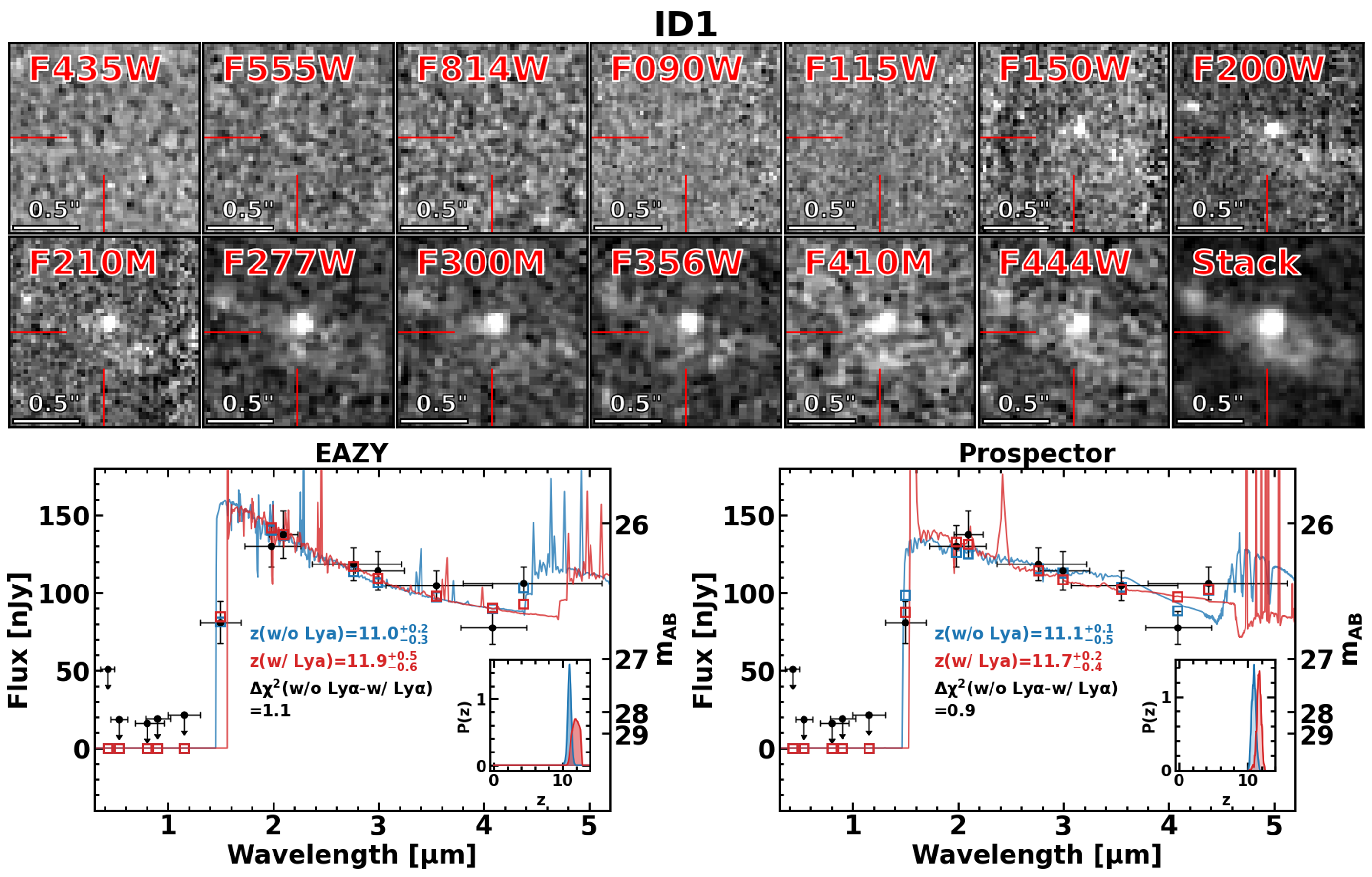}
    \includegraphics[width=0.90\linewidth]{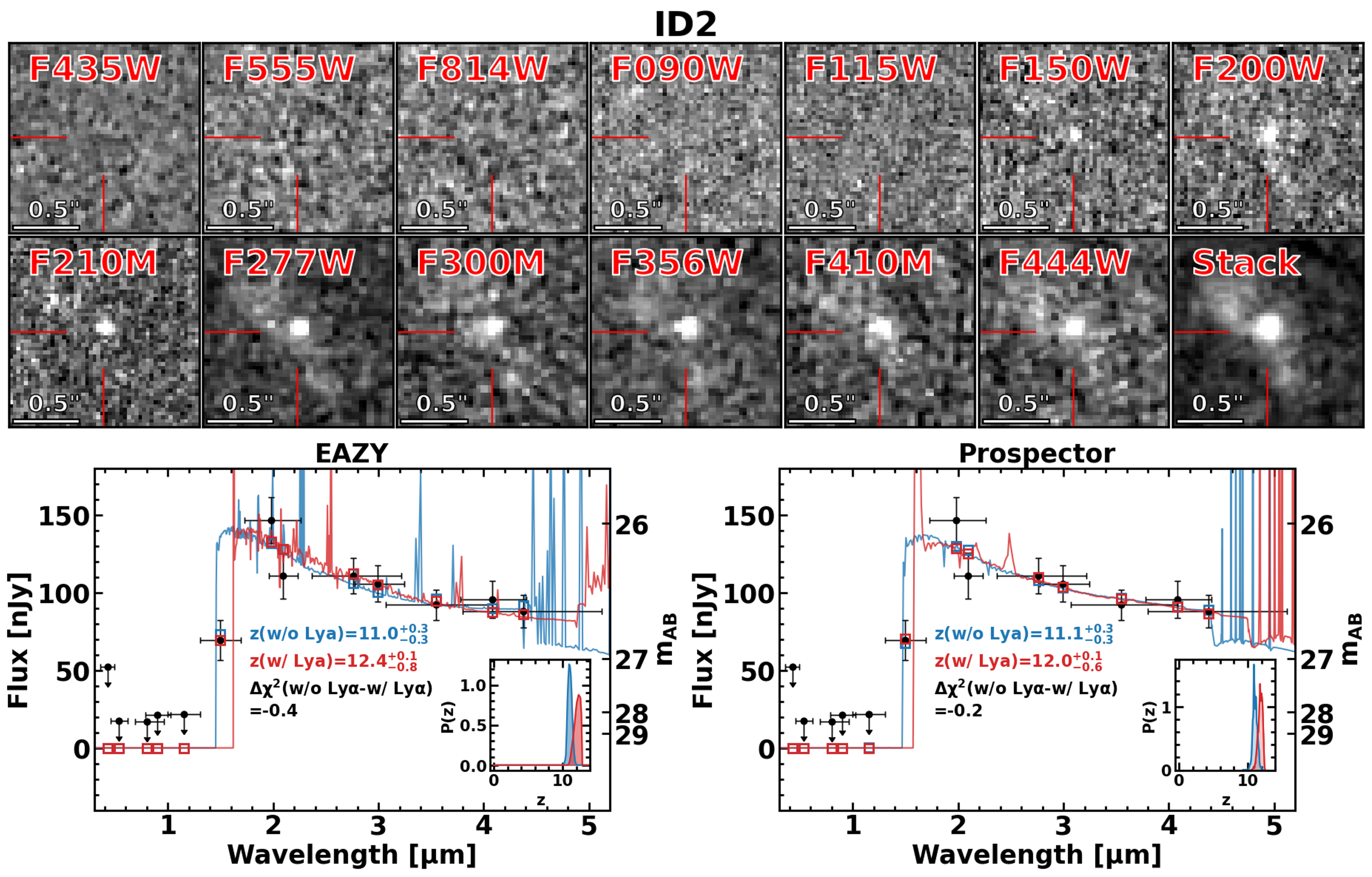}    \caption{JWST + HST images and photometric redshift estimates for ID1 (top) and ID2 (bottom) of the Misty Moons. The upper panels show the \red{$1\farcs5\ \times\ 1\farcs5$} cutout images of $13$ filters (F435W, F555W, F814W, F090W, F115W, F150W, F200W, F210M, F277W, F300M, F356W, F410M, and F444W) and \red{stacked images of F200W, F277W, F356W, and F444W filters}. The lower left and right panels present the results from \texttt{EAZY} and \texttt{Prospector} fitting, respectively. The black circles indicate the photometry measured from the customized apertures with its $1\sigma$ uncertainties (see Section \ref{subsec:photometry}) and wavelength coverages of the filters. For filters with non-detections, the $2\sigma$ upper limits are presented. The fluxes are normalized to total isophotal photometry with the F277W filter. The red curves (squares) indicate best-fit spectra (photometry) with no Ly$\alpha$ emission. The blue curves and squares are the same as the red ones, but with Ly$\alpha$ emission. The inset panels show the posterior probability distributions for redshifts.}
    \label{fig:SED}
\end{figure*}

\begin{figure}[ht!]
    \centering
    \includegraphics[width=0.99\linewidth]{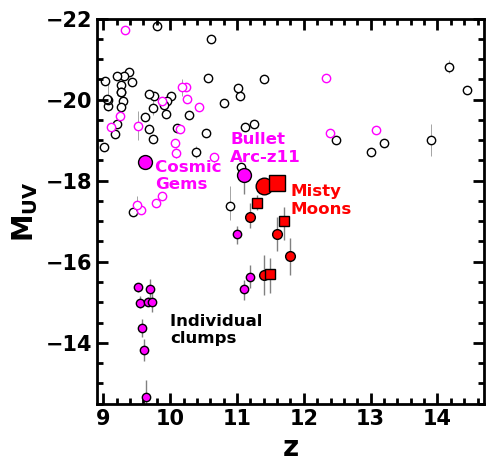}
    \caption{Absolute UV magnitude as a function of redshift. The red filled circles and squares indicate ID1 and ID2 of the Misty Moons ($z_\mathrm{phot}\sim11-12$), respectively. The magenta filled circles represent highly magnified galaxies of the Cosmic Gems arc ($z_\mathrm{spec}=9.63$; \citealt{Adamo2024,Bradley2024,Messa2026}) and BulletArc-z11 ($z_\mathrm{spec}=11.10$; \citealt{Bradac2025}). The small filled symbols show the individual clumps of these galaxies. The magenta and black open circles denote the other spectroscopically confirmed $z>9$ galaxies in lensing fields \citep{Hsiao2023,Hsiao2024,Stiavelli2023,Williams2023,Bradac2024,Fujimoto2024,Marconcini2024,McLeod2024,Roberts-Borsani2024,Napolitano2025,Yanagisawa2025} and blank fields \citep{ArrabalHaro2023,Bunker2023,Curtis-Lake2023,Carniani2024,Castellano2024,D'Eugenio2024,Hainline2024,Harikane2024,Zavala2024,Kokorev2025b,Naidu2025,Napolitano2025,Schouws2025,Tang2025,Witstok2025}, respectively.}
    \label{fig:z-M_UV}
\end{figure}

From these imaging data, we searched for sources with high photometric redshifts and high lensing magnifications. We identified two multiple images of a $z\sim11-12$ galaxy, which exhibit clear F150W dropouts with non-detections in all bands at shorter wavelengths and similar SED colors (Figure \ref{fig:SED}). Based on the independent lens models (see Section \ref{sec:lensmodel}), five multiple images (ID1-ID5) are predicted for this galaxy, two of which, ID1 and ID2, are bright ($\mathrm{F200W}\sim26$ mag) and exhibit complex morphologies owing to their high magnification
($\mu\sim20-30$), as shown in Figure \ref{fig:Misty_Moons}. On the other hand, the intrinsic UV magnitudes derived from the F200W fluxes are as low as $M_\mathrm{UV}\sim-18$ mag (Figure \ref{fig:z-M_UV}). We also find a candidate for ID3, characterized by non-detections at short wavelengths ($<1.5$ $\mu$m) and blue SED just like ID1 and ID2. In the following sections, we mainly focus on the highly magnified multiple images of ID1 and ID2.

\subsection{Photometry}
\label{subsec:photometry}

We produce a detection image by taking an inverse variance-weighted mean of the psf-matched images of F200W, F277W, F356W, and F444W based on the F444W point spread function (PSF). Source detection is conducted on the detection image, using \texttt{Photutils} \citep{Bradley2025}. As shown in Figure \ref{fig:Misty_Moons}, since ID1 and ID2 have clumpy, complex structures, we conduct isophotal photometry in the detection images. We determine isophotal regions by setting two S/N ratio criteria for estimating photometric redshifts and stellar population of the galaxy. \red{For photometric redshift estimates}, to enhance the S/N ratios and get reliable SEDs, we adopt customized apertures, which are determined by regions $6\sigma$ above the background. \red{To accurately estimate the stellar populations of the entire system}, it is required to include faint diffuse fluxes that may come from older stars. We thus determine total isophotal apertures by regions $1.5\sigma$ above the background, \red{which covers the entire system}. We measure the flux of each filter with \texttt{Photutils}, adopting the customized and total isophotal apertures. To measure photometric uncertainties, we place the same customized and isophotal apertures on randomly selected sky regions and measure fluxes of the regions. We derive $1\sigma$ uncertainties from the $68$\% intervals of the resulting flux distributions. The customized aperture fluxes are \red{corrected for total fluxes with a normalization factor obtained from total isophotal fluxes of the F277W filter.} Note that we do not propagate the uncertainty of the \red{normalization factor} because it does not change the colors, which has negligible impacts on the photometric redshift estimates. 
%The isophotal flux of each filter is corrected to the total flux by multiplying a correction factor, which is derived in the following way. We utilize the F277W filter, which shows the highest S/N ratios among the filters, as a reference filter to calculate the correction factor. We measure the total flux of the F277W image with an isophotal aperture, which is determined by a region $2\sigma$ above the background and covers the entire system (Figure \ref{fig:plane}). We then derive the correction factor, dividing the total flux by the small aperture flux. 
For comparison, we also measure aperture photometry, extracted in circular apertures with diameters of $0\farcs3$ and corrected for total values with the Kron apertures \citep{Kron1980}. In addition, the light that is missing beyond the Kron apertures is corrected in a similar method to that used in \citet{Whitaker2011} and \citet{Weaver2024}. The measured photometry is presented in Table \ref{tab:photometry}. The customized aperture fluxes and Kron fluxes are broadly consistent within the uncertainties while the total isophotal fluxes show slightly redder colors probably due to the inclusion of diffuse emission.

We measure UV slopes $\beta$ of ID1 and ID2, using five filters of F200W, F210M, F277W, F300M, and F356W. Since F410M and F444W filters might be impacted by a Balmer break/jump (see Figure \ref{fig:SED}), we do not use them. The results yield relatively blue UV slopes of $\beta=-2.41\pm0.20$ for ID1 and $\beta=-2.60\pm0.23$ for ID2.

\subsection{Photometric Redshift}
\label{subsec:photo-z}
To derive photometric redshifts, we perform SED fitting to the customized aperture fluxes of ID1 and ID2 with \texttt{EAZY} \citep[ver. 0.8.5; ][]{Brammer2008}. We fit the blue SED templates of \citet{Larson2023} which are based on stellar population models of BPASS \citep{Eldridge2017,Stanway2018} and photoionization models of \texttt{Cloudy} \citep{Ferland2017}, and reflects blue UV slopes of observed galaxies at $z>8$. We perform two sets of fitting, either including or excluding Ly$\alpha$ emission. For the models including Ly$\alpha$ emission, we apply absorption models of the intergalactic medium (IGM; \citealt{Inoue2014}). \red{For the models without Ly$\alpha$ emission, we consider the additional Ly$\alpha$ damping wing absorption arising from the dense} H \textsc{i} \red{gas clouds in the circumgalactic medium (CGM), adopting the IGM + CGM absorption models \citep{Asada2025}.} While the hydrogen in the IGM is mostly neutral at $z>10$, Ly$\alpha$ emission could still be observable if the galaxy hosts strong ionizing sources, such as faint systems with blue UV slopes that efficiently produce and leak ionizing photons \citep[e.g., ][]{Witstok2025}. We thus conduct SED fitting, considering the possible contribution of Ly$\alpha$ emission. 
We also perform \texttt{Prospector} \citep{Johnson2021} SED fitting on the same photometry, as done with \texttt{EAZY}. We assume simple stellar populations (SSPs) with Flexible Stellar Population Synthesis \citep[FSPS; ][]{Conroy2010} where we adopt the MIST isochrones \citep{Choi2016} and MILES stellar library \citep{Sanchez-Blazquez2006}. Following \citet{Wang2024}, the composite stellar populations (CSPs) are constructed with \texttt{Prospector-$\beta$} \citep{Wang2023}, where we assume the initial mass function (IMF) of \citet{Chabrier2003}, a two component dust model \citep{Charlot2000}, and IGM absorption models of \citet{Madau1995}. As in \texttt{EAZY}, we either include or exclude Ly$\alpha$ emission. We adopt a flexible non-parametric star formation history (SFH) with seven logarithmically spaced time bins and a dynamic SFH prior, which depends on the redshift and stellar mass, and reflects the cosmic SFR density \citep{Behroozi2019}. See \citealt{Wang2023,Wang2024} for further details of the prior.

In Figure \ref{fig:SED}, we present the fitting results. The best-fit models excluding Ly$\alpha$ emission for both ID1 and ID2 yield redshifts of $z\sim11$ while those including Ly$\alpha$ emission shifts the redshifts to $z\sim12$. From the $\Delta\chi^2$ value for each fitting (see Figure \ref{fig:SED}), we cannot exclude either model including or excluding Ly$\alpha$ emission. In any case, the blue SEDs and non-detections in short-wavelength filters at $<1.5$ $\mu$m support a high probability of $P(z>10)\sim1.0$. We present the redshift and absolute UV magnitude distribution of $z>9$ galaxies in Figure \ref{fig:z-M_UV}.
%While the best-fit models of ID1 (ID2) from \texttt{EAZY} suggest the redshift to be $z_\mathrm{phot}=10.9_{-0.2}^{+0.3}$ ($11.3_{-0.5}^{+0.4}$), those from \texttt{Prospector} indicate the redshift to be $z_\mathrm{phot}=11.9_{-0.3}^{+0.3}$ ($12.1_{-0.3}^{+0.2}$) with a sharp Ly$\alpha$ break or Ly$\alpha$ emission. Since the central clumps of ID1 and ID2 exhibit high surface density (Section \ref{subsec:property}), there could be a strong UV radiation field, which may allow Ly$\alpha$ emission to remain detectable. Therefore, the Ly$\alpha$ emission is not necessarily weak and the \texttt{Prospector} best-fit SEDs may still be valid. Due to the strong lensing, ID1 and ID2 are resolved into multiple clumps. We describe the lens models and magnification factors in the next section.

\begin{table*}[ht!]
    \centering
    \footnotesize
    \caption{Properties of the Misty Moons.}
    \begin{tabular}{ccccccccc}
    \hline
    \hline
    Name & R.A. & Decl. & $z_\mathrm{\texttt{EAZY}}$ & $z_{\mathrm{\texttt{EAZY}}+\mathrm{Ly}\alpha}$& $z_\mathrm{\texttt{Prospector}}$& $z_{\mathrm{\texttt{Prospector}}+\mathrm{Ly}\alpha}$ & $M_\mathrm{UV}$&$\beta$\\
    &(deg)&(deg)&&&&&&\\
    \hline
         ID1 &-&-& $11.0_{-0.3}^{+0.2}$ & $11.9_{-0.6}^{+0.5}$ & $11.1_{-0.3}^{+0.3}$ & $12.0_{-0.6}^{+0.1}$ & $-17.90_{-0.15}^{+0.15}$ & $-2.41\pm0.20$\\
         A.1 &$44.287102$&$-23.437968$&-&-&-&-& $-17.12_{-0.33}^{+0.28}$ & $-2.60\pm0.18$\\
         B.1&$44.287275$&$-23.437903$&-&-&-&-& $-15.66_{-0.39}^{+0.35}$ & $-1.83\pm0.53$\\
         C.1&$44.287029$&$-23.438028$&-&-&-&-& $-16.70_{-0.41}^{+0.32}$ & $-3.47\pm0.51$\\
         D.1&$44.286927$&$-23.438012$&-&-&-&-& $-16.13_{-0.42}^{+0.43}$ & $-2.71\pm0.76$\\
         ID2&-&-& $11.0_{-0.3}^{+0.3}$ & $12.4_{-0.8}^{+0.1}$ & $11.1_{-0.3}^{+0.3}$ & $12.0_{-0.6}^{+0.1}$ & $-17.93_{-0.19}^{+0.18}$ & $-2.60\pm0.23$\\
         A.2&$44.288122$&$-23.437200$&-&-&-&-& $-17.46_{-0.17}^{+0.15}$ & $-2.94\pm0.15$ \\
         B.2&$44.288080$&$-23.437306$&-&-&-&-& $-15.70_{-0.38}^{+0.47}$ & $-1.99\pm0.77$\\ 
         C.2&$44.288215$&$-23.437158$&-&-&-&-& $-17.00_{-0.25}^{+0.26}$ &$-2.80\pm0.61$ \\
         \hline
    \end{tabular}
    \begin{tablenotes}  
    \footnotesize
    \item Notes. Photometric redshifts are measured for customized aperture fluxes of ID1 and ID2, using \texttt{EAZY} and \texttt{Prospector}. We present two types of redshifts, one of which is based on SED models without Ly$\alpha$ emission ($z_\mathrm{\texttt{EAZY}}$ and $z_\mathrm{\texttt{Prospector}}$), while the other is based on SED models with Ly$\alpha$ emission ($z_{\mathrm{\texttt{EAZY}}+\mathrm{Ly}\alpha}$ and $z_{\mathrm{\texttt{Prospector}}+\mathrm{Ly}\alpha}$). The $M_\mathrm{UV}$ values are corrected for magnification.
    \end{tablenotes}
    \label{tab:property1}
\end{table*}

\begin{table*}[ht!]
    \centering
    \caption{Properties of the Misty Moons (continued).}
    \begin{tabular}{cccccccc}
    \hline
    \hline
    Name &$\log(M_\mathrm{*,int})$&$r_\mathrm{eff}$ &$r_\mathrm{eff,int}$ & $\log(\Sigma_*)$ & $\log(\Sigma_\mathrm{SFR})$& $\mu_\mathrm{Zitrin-Analytic}$ & $\mu_\mathrm{Glafic}$\\
    &($M_\odot$)&(pc)&(pc)&($M_\odot\ \mathrm{pc^{-2}}$)&($M_\odot\ \mathrm{yr^{-1}}\ \mathrm{kpc^{-2}}$)&\\
    \hline
         %ID1& $7.65\pm0.21$ & $958\pm78$& $181\pm18$ & $2.39\pm0.22$ &$0.36\pm0.13$&$31.0_{-1.0}^{+8.0}$ & $28.0_{-3.0}^{+3.0}$\\
         ID1& $7.57\pm0.26$ & $958\pm78$& $181\pm18$ & $2.25\pm0.27$ &$0.46\pm0.11$&$31.0_{-1.0}^{+8.0}$ & $28.0_{-3.0}^{+3.0}$\\
         A.1& $7.03\pm0.19$ & $83\pm11^\mathrm{a}$ & $15\pm3^\mathrm{a}$ & $3.88\pm0.19^\mathrm{a}$ & $2.24\pm0.13^\mathrm{a}$ & $30.6_{-6.7}^{+1.8}$ & $30.8_{-7.8}^{+8.1}$\\
         &&$<203^\mathrm{b}$ & $<20^\mathrm{b}$ & $>3.62^\mathrm{b}$ & $>1.97^\mathrm{b}$&&\\
         B.1& $7.09\pm0.41$ & $134\pm24$ & $22\pm5$ & $3.59\pm0.41$ & $1.83\pm0.37$&$29.6_{-6.7}^{+1.6}$ & $37.4_{-10.3}^{+11.2}$\\
         C.1& $6.87\pm0.24$ & $347\pm34$ & $67\pm12$ & $2.40\pm0.23$ & $0.66\pm0.16$&$33.9_{-8.4}^{+1.6}$ & $27.0_{-6.4}^{+8.1}$\\
         D.1& $6.92\pm0.36$ & $189\pm39$ & $36\pm8$ & $3.02\pm0.38$ & $1.19\pm0.33$&$40.6_{-11.2}^{+2.9}$ & $27.4_{-6.4}^{+6.8}$\\
         %ID2& $7.50\pm0.12$ & $903\pm116$ & $197\pm28$ & $2.15\pm0.16$ & $0.43\pm0.13$&$35.0_{-9.0}^{+2.0}$ & $21.0_{-3.0}^{+3.0}$ \\
         ID2& $7.53\pm0.22$ & $903\pm116$ & $197\pm28$ & $2.15\pm0.24$ & $0.45\pm0.14$&$35.0_{-9.0}^{+2.0}$ & $21.0_{-3.0}^{+3.0}$ \\
         A.2& $7.02\pm0.14$ & $83\pm42^\mathrm{a}$ & $17\pm9^\mathrm{a}$ & $3.72\pm0.51^\mathrm{a}$ & $2.18\pm0.47^\mathrm{a}$ & $31.0_{-0.6}^{+7.2}$ & $21.4_{-3.0}^{+2.8}$ \\
         &&$<203^\mathrm{b}$ & $<26.4^\mathrm{b}$ & $>3.38^\mathrm{b}$ & $>1.84^\mathrm{b}$&&\\
         B.2& $7.11\pm0.35$ & $147\pm145$ & $36\pm33$ & $3.24\pm0.86$ & $1.28\pm0.85$&$40.2_{-0.9}^{+14.1}$ & $28.3_{-3.9}^{+3.7}$\\ 
         C.2& $6.98\pm0.24$ & $296\pm111$ & $56\pm21$ & $2.74\pm0.46$ & $0.98\pm0.39$&$26.6_{-0.5}^{+4.9}$ & $18.6_{-2.4}^{+2.5}$\\         
         \hline
    \end{tabular}
    \begin{tablenotes}  
    \footnotesize
    \item Notes. The $M_\mathrm{*,int}$ and $\mathrm{r_{eff,int}}$ values are corrected for magnification while the $\mathrm{r_{eff}}$ values are measured in the image plane. We derive $\Sigma_\mathrm{SFR}$ values, using the SFRs in the recent $10$ Myr derived from \texttt{Prospector} SED fitting. \red{$^\mathrm{a}$ The size and surface density measurements from detection images where these clumps are marginally resolved. $^\mathrm{b}$ The size and surface density limits from F200W images where these clumps are unresolved (see Section \ref{subsec:clump}).} 
    \end{tablenotes}
    \label{tab:property2}
\end{table*}

\section{Lens Models} 
\label{sec:lensmodel}
In this work, we construct two independent lens models, one with a revised version of the \citet{Zitrin2015} parametric method (a.k.a Zitrin-Analytic; see \citealt{Furtak2024} for more details), and Glafic \citep{Oguri2010,Oguri2021}. These lens models are constructed based on previously known multiple images (e.g. \citealt{Zitrin2011}) together with new identifications with NIRCam in the VENUS program, as shown in the colored circles in Figure \ref{fig:Misty_Moons}. The predictions of the five multiple images (ID1-ID5) and their magnification values are broadly consistent among the two independent models. As described in Section \ref{subsec:photometry}, we identify two highly magnified counter images (ID1 and ID2) and a candidate for ID3. We do not detect ID4 and ID5, which is consistent with our model predictions of their low magnifications. We estimate the expected apparent magnitudes in F200W for ID4 and ID5 from that of ID1 and their magnification factors (Sections \ref{subsec:PIEMDeNFW} and \ref{subsec:Glafic}). The results are $\mathrm{F200W}\sim29$ mag, which is $\sim1$ mag fainter than the $5\sigma$ detection limit of our NIRCam observations ($\sim28$ mag; Section \ref{subsec:obs}). Further deep NIRCam imaging would be required to identify ID4 and ID5. We note that the geometry predicted by two independent lens models is similar to a hyperbolic umbilic (H-U) lensing configuration (e.g., \citealt{Limousin2008,Meena2020,Lagattuta2023}). This suggests that if all the five multiple images are confirmed, the presence of the H-U configuration could provide strong constraints on the radial profile of the lens. We detail the two models in the following sections.

\subsection{Zitrin-Analytic (dPIE-PIEMD) Model}
\label{subsec:PIEMDeNFW}
\begin{figure}[ht!]
    \centering
    \includegraphics[width=0.99\linewidth]{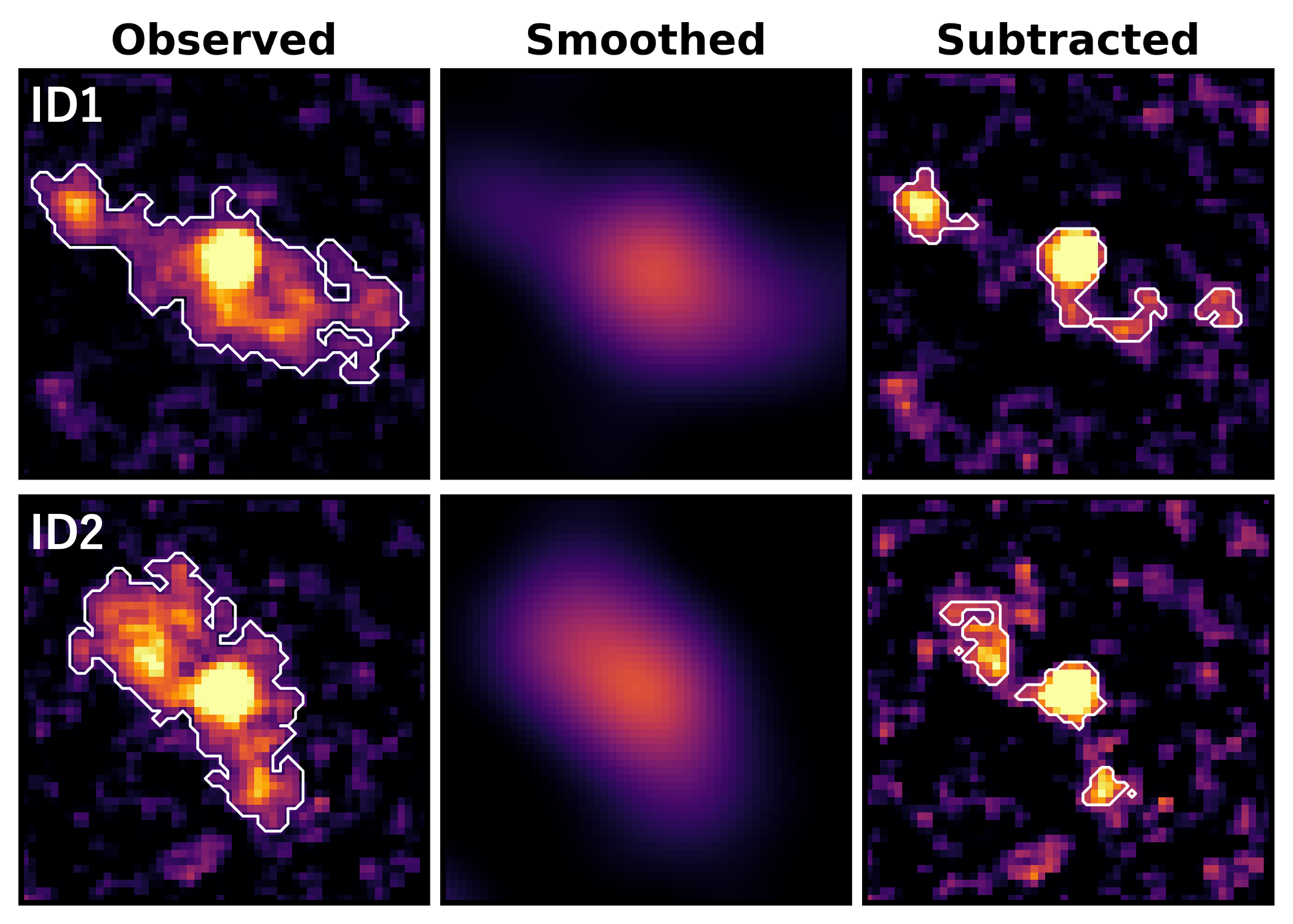}
    \caption{\red{Clump detection procedure for ID1 and ID2. From left to right, each panel presents the observed detection images (F200W + F277W + F356 W + F444W), smoothed-host images, and host-subtracted contrast images (Section \ref{subsec:detection}). The white contours in the left and right panels indicate isophotal regions detected with \texttt{Photutils}.}}
    \label{fig:clump_detection}
\end{figure}

In this model we include 10 sets of multiply imaged systems, including multiple knots in the three multiply imaged systems $1,2,3$ from \citet{Zitrin2011}; the other less secure systems found by \citet{Zitrin2011} seem not to be true multiple image associations. Using MUSE data we determine the redshift of all three images of system $1$ to be $z_\mathrm{spec}=1.09$, which also spectroscopically confirms that all images are of the same system. We identify four new systems in JWST VENUS data, using also public data obtained by the SLICE program (PID: 5594; PI: Mahler): systems $(6,7,8)$ and the Misty Moons system. We do not include the third multiple image candidate (ID3) for this system and only use the two clearly identified images (ID1\&ID2) to constrain the model. Given its dropout nature, we set the redshift of the Misty Moons to be $z=12$ \footnote{Note that at this high source redshift the \emph{exact} redshift has a negligible effect of the relative Dls/Ds ratio and thus, model.}, and system 1 to $z=1.09$ and leave the redshifts of the other systems to be optimized by the model. The lens model is constructed using the so-called Zitrin-Analytic technique which is a revised version of the \citet{Zitrin2015} parametric method (see also \citealt{Pascale2022,Furtak2023} for more details). Cluster members are modeled as double Pseudo Isothermal Ellipsoids (dPIE; \citealt{Elıasdottir2007}) and the dark matter component of the cluster mass distribution is treated as a sum of diffused halos, modeled as Pseudo Isothermal Elliptical Mass Distributions (PIEMDs; e.g. \citealt{Keeton2001}). We find that one main dark matter halo centered on the main BCG but with the exact position free to move, is sufficient to well reproduce the lensing features in this cluster, but also add a second, smaller halo centered on the second BCG to improve its accuracy. We do not include here an external shear. The model reproduces all multiple images, with a $\chi^2$ of $9.4$ with 11 degrees-of-freedom (i.e. a reduced $\chi^2\simeq0.9$), and an rms of $0\farcs28$. For the minimization we first adopt a positional uncertainty of $0\farcs3$, but for the final error and $\chi^2$ calculation we adopt the more commonly used positional uncertainty of $0\farcs5$. For the Misty Moons, the model predicts three more images at RA,Dec : ($44.28487483$, $-23.43738994$), ($44.28650147$, $-23.43596856$), ($44.28569941$, $-23.42626713$).
The calculated magnifications for ID1 and ID2 are $31^{+8}_{-1}$ and $35^{+2}_{-9}$, and for the other three predicted images they are $30.4^{+3.6}_{-0.8}$, $6.9^{+1.6}_{-0.5}$, and $2.1^{+0.1}_{-0.1}$, respectively. The third image at (RA, Dec)=($44.28487483$, $-23.43738994$) is a bit apart from the candidate for ID3, shown in Figure \ref{fig:Misty_Moons}, and the magnification factor is thus deviated from that for the candidate of ID3 predicted by the Glafic model (Section \ref{subsec:Glafic}). The magnification derived with the Zitrin-Analytic model at the position of the candidate of ID3 is $12.7^{+2.8}_{-0.2}$, which is comparable to that estimated with the Glafic model (Section \ref{subsec:Glafic}).

\subsection{Glafic Model}
\label{subsec:Glafic}
\begin{figure}[ht!]
    \centering
    \includegraphics[width=0.99\linewidth]{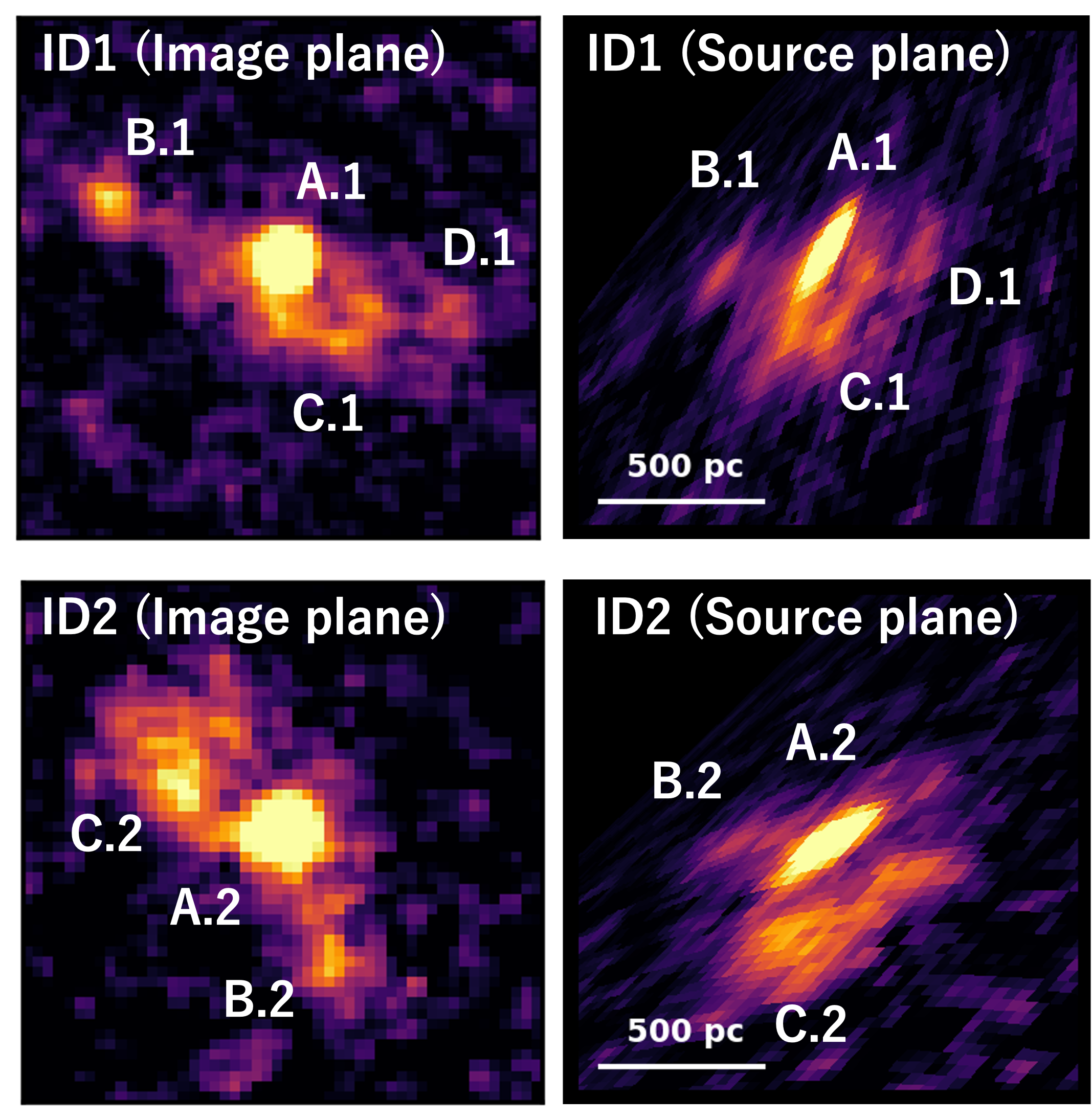}
    \caption{Detection images (F200W + F277W + F356W + 444W) of ID1 and ID2 in the image plane (left) and source plane (right).}
    \label{fig:plane}
\end{figure}

We include three multiple image systems identified in \citet{Zitrin2011} as well as three additional multiple image systems identified in the VENUS data including the Misty Moons as observational constraints. For the Misty Moons, in addition to ID1 and ID2, the position of the ID3 candidate at ($44.2845714$, $-23.4367132$) is also included. We model the mass distribution assuming a Navarro-Frenk-White (NFW; \citealt{Navarro1996}) profile for the halo component and member galaxy perturbations modeled by pseudo-Jaffe ellipsoids. In addition, the external shear is included. Since no spectroscopic redshift is available for all these multiple image systems except system $1$, we include their photometric redshift as priors. Assuming the positional error of $0\farcs4$, our best-fitting model has $\chi^2=14.7$ for the degree of freedom of $15$. The rms between model-predicted and observed multiple image positions is $0\farcs36$. The model predicts the magnification of ID1 of $28\pm3$, ID2 of $21\pm3$, and ID3 of $11\pm 1$. It also predicts additional images (ID4, ID5) at around (44.2862877, $-23.4354116$) and (44.2848497, $-23.4255854$), with their magnifications of $2.4\pm 1.3$ and $2.6\pm0.2$, respectively. Their low magnifications well explain the non-detection in the VENUS data.

\begin{figure*}[ht!]
    \centering
    \includegraphics[width=0.95\linewidth]{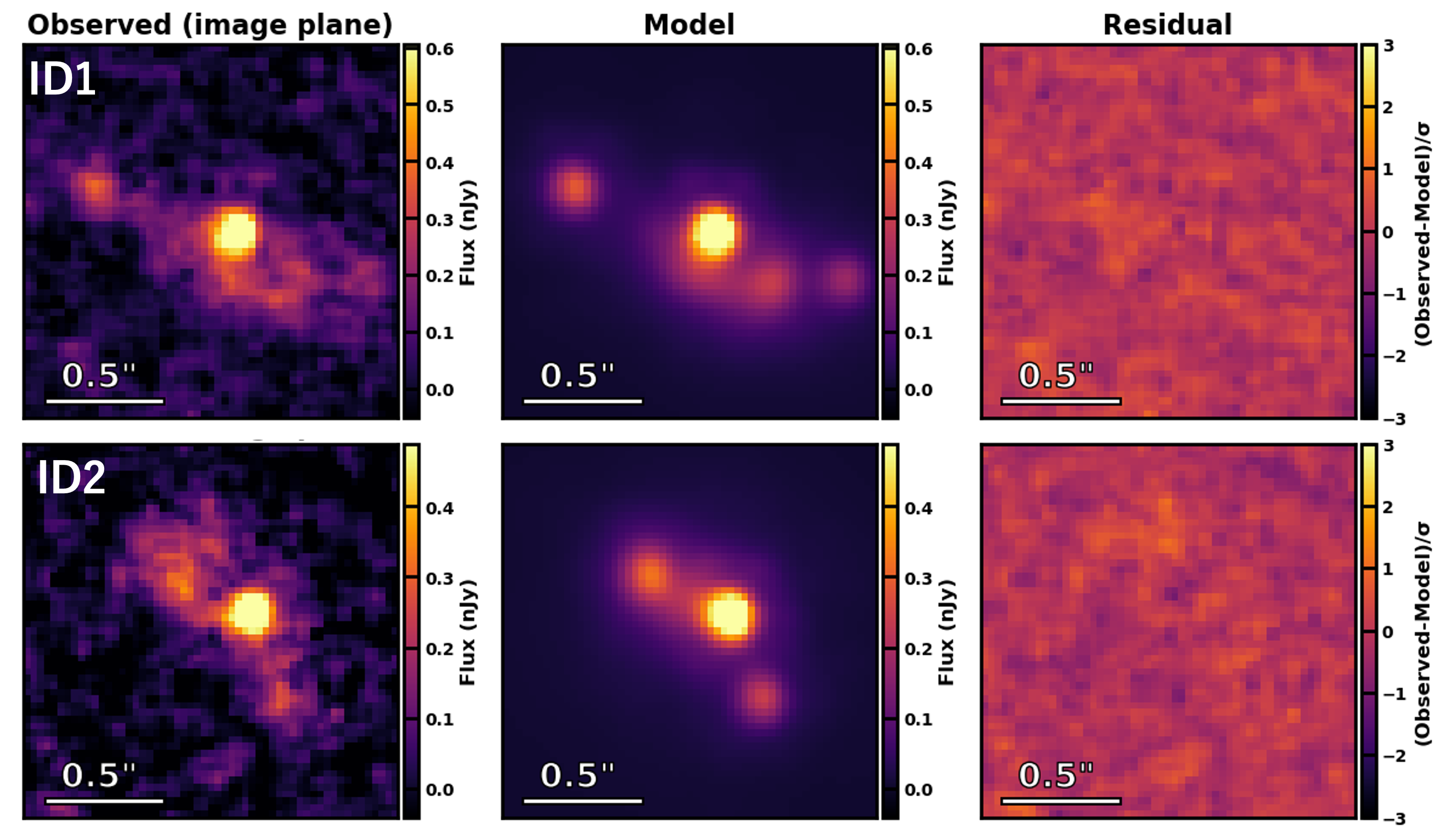}
    \caption{Clump modeling results for ID1 and ID2 of the Misty Moons system. From left to right, the $1\farcs5\times1\farcs5$ observed images, best-fit model images, and residual images divided by the $1\sigma$ error maps are presented.}
    \label{fig:clump_modeling}
\end{figure*}
\section{Analysis}
\label{sec:analysis}

\subsection{Clump Detection}
\label{subsec:detection}
We conduct clump identification, following previous studies \citep[e.g., ][]{Conselice2003,Calabro2019,Kalita2024}. We use the detection image constructed from F200W, F277W, F356W, and F444W filter images (see Section \ref{subsec:photometry}) and smooth it, using Gaussian kernel with a standard deviation of $\sigma=6$ pixel. The smoothed image is subtracted from the original image, leaving behind a contrast map. We perform source extraction in the contrast map with a threshold of $\mathrm{S/N}=2$, using \texttt{Photutils} \citep{Bradley2025}, and identify four clumps (A.1, B.1, C.1, and D.1) for ID1 and three clumps (A.2, B.2, and C.2) for ID2 (see Figure \ref{fig:clump_detection}), which are detected in almost all filters at $>1.5$ $\mu$m. \red{In Figure \ref{fig:plane}, we compare image-plane and source-plane images of ID1 and ID2. The source plane images are reconstructed pixel-by-pixel with the Glafic model.} Despite the lensing distortions, the reconstructed morphologies and relative positions are highly similar, suggesting that the individual clumps of ID1 and ID2 may correspond to the same intrinsic structures.

\subsection{Clump Modeling}
\label{subsec:clump}
To simultaneously model the multiple clumps, we conduct S\'ersic profile fitting using \texttt{GALFIT} \citep{Peng2010}.
%We measure the half-light radius $r_\mathrm{eff}$.
For ID1, we adopt a five-component model consisting of five S\'ersic profiles to simultaneously fit the four individual clumps and diffuse light. In the similar way, a four-component model consisting of four S\'ersic profiles is used for ID2. We set the initial parameters of the fitting with the clump identification results with \texttt{Photutils} described in the previous section. We fix the S\'ersic index to $n=1$ for the individual clumps except for the diffuse light, and set the boundary limits of the other parameters in the same manner as \citet{Fujimoto2025}. \red{We fit to the detection images because of their higher S/N ratios than the individual filter images.} The best-fit models and residuals of ID1 and ID2 are presented in Figure \ref{fig:clump_modeling}. The intrinsic half-light radius in the source plane, $r_\mathrm{eff,int}$, is obtained by dividing the measured half-light radius in the image plane by the square root of the magnification factor predicted by the Glafic model. \red{The most compact clumps of A.1 and A.2 are only marginally larger than the PSF in the detection images, leaving modest residuals after pure PSF fitting. To further investigate whether these clumps are spatially resolved, we additionally analyze the F200W images which have a smaller PSF and therefore provide higher spatial resolution. In the F200W images, the clumps are well reproduced by the PSF without significant residuals, suggesting that they may be unresolved. The similarly point-like appearance of these clumps across multiple images is also consistent with the unresolved interpretation. The corresponding radial surface-brightness profiles are presented in Figure \ref{fig:sb_profile}. If these clumps are unresolved, we adopt half width at half maximum of the PSF ($0\farcs05$) as an upper limit of the observed effective radius. The corresponding upper limit of the intrinsic effective radius is estimated by dividing the observed one by the tangential magnification ($\mu_\mathrm{tan}=10.1$ for A.1 and $7.7$ for A.2). The apparent discrepancy between the detection and F200W images can be interpreted as follows. If these clumps are marginally resolved, the low S/N ratio of the F200W images may prevent detection of the low-surface-brightness outskirts of the clumps, causing them to appear PSF-like. Alternatively, if these clumps are unresolved, the slightly extended appearance in the detection images may arise from residual contamination by the diffuse component and correlated noise. Because the current data do not allow us to distinguish between these two scenarios, we report both the best-fit effective radii assuming marginally resolved clumps and upper limits assuming unresolved clumps in Table \ref{tab:property2}.} 

We measure the clump fluxes by fitting the same models as used for the $r_\mathrm{eff}$ measurement to the PSF-matched filter images. We set only the fluxes as free parameters, and fix the other parameters of the S\'ersic profiles to the best-fit ones obtained from the detection images. To derive the flux uncertainties, we conduct Monte Carlo simulations. We construct mock observation images by injecting the PSF convolved best-fit models to randomly selected sky regions with no other sources. We perform identical flux measurements with the mock observation images, and determine the $1\sigma$ uncertainties from the $68$\% intervals of the resulting flux distributions. 

\subsection{SED Fitting}
\label{subsec:sed}
To derive stellar populations of the entire systems and all clumps of ID1 and ID2, we conduct SED fitting using \texttt{Prospector} with flexible non-parametric SFHs in the same way as described in Section \ref{subsec:photo-z}. To focus on the stellar population properties, we fix the redshift to be $z=11.5$, which is a mean value of all the photometric redshifts presented in Table \ref{tab:property1}. We use total isophotal fluxes for the entire systems of ID1 and ID2. For comparison of clump stellar populations, we also perform \texttt{Bagpipes} \citep{Carnall2018} SED fitting with short SFHs, which is conducted for high-$z$ star-forming clumps \citep{Vanzella2022,Adamo2024,Bradley2024,Claeyssens2025,Fujimoto2025}. We adopt SPS models by \citet{Bruzual2003} and include nebular emission calculated with \texttt{Cloudy}. We fix the redshift to be $z=11.5$, same as \texttt{Prospector} fitting, and assume a \citet{Calzetti2000} attenuation law. For SFHs, we assume delayed-$\tau$ models, expressed as $\mathrm{SFR}\propto te^{-t/\tau}$, where we adopt a short timescale of $\tau=10$ Myr, following the studies of high-$z$ clumps/star clusters (e.g., \citealt{Adamo2024,Claeyssens2025}). We adopt flat priors for formed mass $5<\log(M_*/M_\odot)<13$, formed age $1\ \mathrm{Myr}<t<1\ \mathrm{Gyr}$, formed metallicity $0.01<Z_*/Z_\odot<1$, V-band extinction $0<A_\mathrm{V}<3$, and ionization parameter $-3<\log(U)<-1$. In Table \ref{tab:sed}, we present the inferred properties from the SED fitting.

\begin{figure*}[ht]
    \centering
    \includegraphics[width=0.95\linewidth]{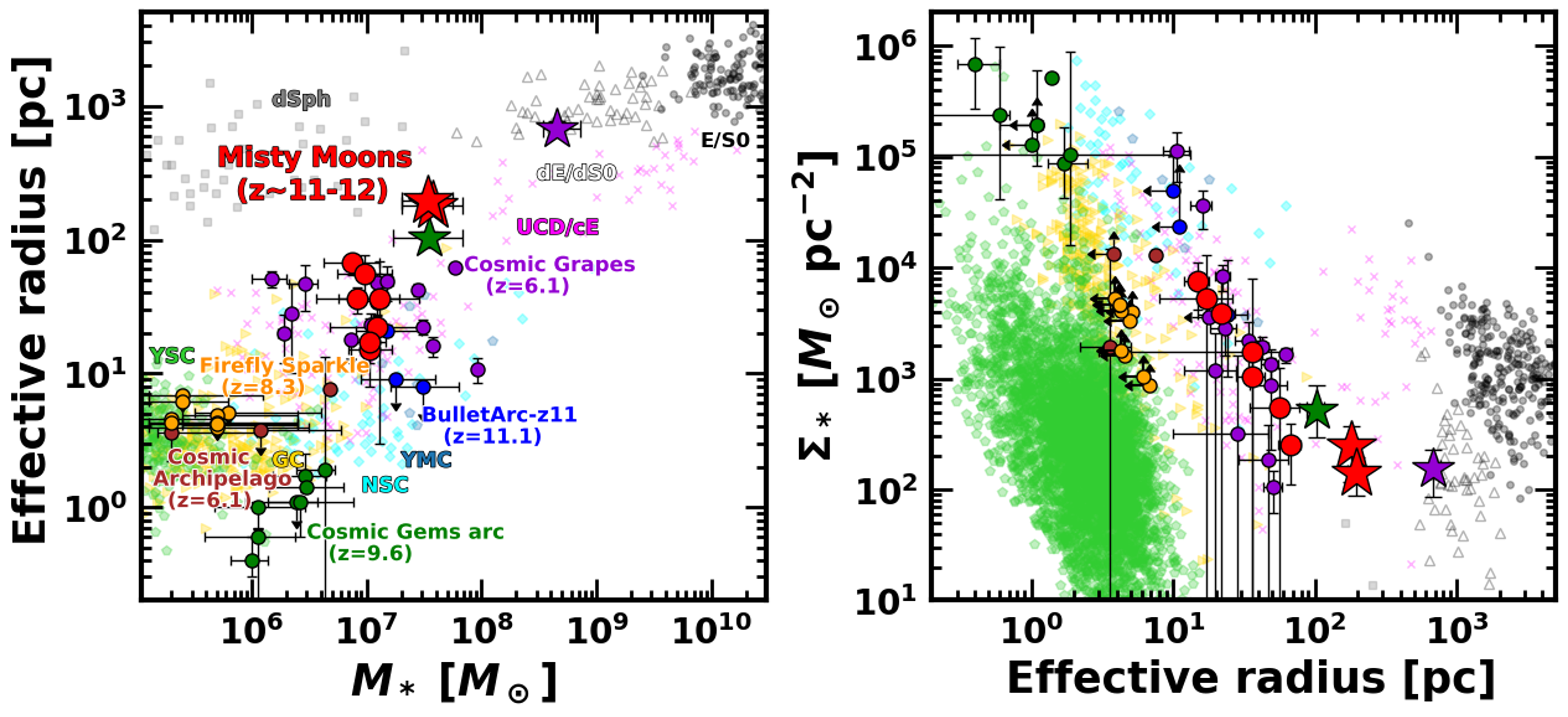}
    \caption{Physical properties of individual clumps. Left: stellar mass versus effective radius relation. Right: stellar mass surface density as a function of effective radius. The star symbols indicate the entire systems while the circles are individual clumps. The red, blue, green, orange, brown, and purple symbols show the Misty Moons ($z_\mathrm{phot}\sim11-12$; This work), BulletArc-z11 ($z_\mathrm{spec}11.10$; \citealt{Bradac2025}), Cosmic Gems arc ($z_\mathrm{spec}=9.63$; \citealt{Adamo2024,Bradley2024,Messa2026,Vanzella2026}, Firefly Sparkle ($z_\mathrm{spec}=8.27$; \citealt{Mowla2024}), Cosmic Archipelago ($z_\mathrm{spec}=6.14$; \citealt{Messa2025}), and Cosmic Grapes ($z_\mathrm{spec}=6.07$; \citealt{Fujimoto2025}), respectively. For comparison, the other symbols represent local galaxies or star clusters of elliptical/S0 galaxies (E/S0; gray circles), dwarf elliptic/S0 galaxies (dEs/dS0; open triangles), dwarf spheroids (dSph; gray squares), ultra-compact dwarfs/compact elliptical galaxies (UCD/cE; magenta crosses), young massive star clusters (YMC; blue pentagon), nuclear star clusters (NSCs; cyan diamonds), globular clusters (GC; yellow triangles), young star clusters (YSC; green pentagon) taken from the compilation in \citet{Norris2014}.}
    \label{fig:size_density}
\end{figure*}

\begin{table*}[ht!]
    \centering
    \caption{Physical properties of the Misty Moons in the image plane estimated from SED fitting.}
    \begin{tabular}{cccccc}
    \hline
    \multicolumn{6}{c}{\texttt{Prospector}, flexible non-parametric SFH} \\ 
    \hline
    Name & $\log(\mu M_*)$ & Age & A$_\mathrm{V}$&$\mu$SFR ($10$ Myr) &$\mu$SFR ($100$ Myr) \\
    &($M_\odot$)&(Myr)&(mag)&($M_\odot\ \mathrm{yr^{-1}}$)&($M_\odot\ \mathrm{yr^{-1}}$)\\
    \hline
         %ID1 & $9.13_{-0.23}^{+0.18}$ & $97.1_{-30.2}^{+17.6}$& $0.02_{-0.01}^{+0.03}$& $13.3_{-3.4}^{+3.3}$& $10.2_{-3.0}^{+3.8}$\\
         ID1 & $9.01_{-0.24}^{+0.19}$ & $62.1_{-28.0}^{+41.7}$& $0.02_{-0.02}^{+0.04}$& $16.9_{-3.4}^{+3.3}$& $9.0_{-2.8}^{+2.0}$\\
         A.1 & $8.50_{-0.15}^{+0.15}$ & $42.4_{-19.2}^{+26.5}$& $0.02_{-0.01}^{+0.03}$& $7.4_{-1.0}^{+1.2}$& $3.2_{-0.9}^{+1.2}$\\
         B.1& $8.64_{-0.33}^{+0.44}$ & $56.1_{-27.3}^{+44.2}$& $0.24_{-0.16}^{+0.49}$& $6.8_{-2.9}^{+9.7}$& $3.9_{-1.9}^{+6.7}$\\
         C.1& $8.28_{-0.19}^{+0.25}$ & $54.2_{-26.2}^{+34.2}$& $0.03_{-0.02}^{+0.06}$& $3.6_{-0.8}^{+1.1}$& $1.9_{-0.6}^{+1.1}$\\
         D.1& $8.36_{-0.27}^{+0.42}$& $66.7_{-33.4}^{+44.1}$& $0.14_{-0.10}^{+0.27}$ & $3.4_{-1.2}^{+3.0}$& $2.1_{-0.9}^{+2.4}$\\
         %ID2 & $8.85_{-0.11}^{+0.13}$ & $48.2_{-17.5}^{+23.0}$& $0.02_{-0.02}^{+0.03}$& $13.7_{-1.5}^{+1.8}$& $7.2_{-1.4}^{+2.0}$\\
         ID2 & $8.86_{-0.19}^{+0.22}$ & $58.5_{-28.2}^{+40.0}$& $0.02_{-0.02}^{+0.03}$& $14.5_{-2.4}^{+2.4}$& $6.8_{-2.0}^{+2.5}$\\
         A.2& $8.35_{-0.10}^{+0.16}$ & $35.9_{-14.5}^{+21.7}$& $0.01_{-0.00}^{+0.01}$& $6.5_{-0.6}^{+0.6}$& $2.4_{-0.5}^{+0.8}$\\
         B.2& $8.34_{-0.3}^{+0.4}$ & $62.6_{-30.1}^{+47.5}$& $0.14_{-0.10}^{+0.25}$& $3.1_{-1.1}^{+2.6}$& $1.9_{-0.8}^{+2.3}$\\
         C.2& $8.45_{-0.22}^{+0.32}$& $64.7_{-33.2}^{+47.0}$& $0.06_{-0.05}^{+0.12}$ & $4.9_{-1.3}^{+1.8}$& $2.6_{-0.9}^{+1.8}$\\
    \hline
    \multicolumn{6}{c}{\texttt{Bagpipes}, delayed-$\tau$ ($\tau=10$ Myr) SFH} \\ 
    \hline
    Name & $\log(\mu M_*)$ & Age &A$_\mathrm{V}$&$\mu$SFR & \\
    &($M_\odot$)&(Myr)&(mag)&($M_\odot\ \mathrm{yr^{-1}}$)&\\
    \hline
         A.1 & $8.51_{-0.30}^{+0.36}$ & $15.6_{-9.7}^{+15.0}$& $0.06_{-0.04}^{+0.08}$& $3.8_{-2.1}^{+5.4}$& \\
         B.1& $9.05_{-0.59}^{+0.32}$ & $42.6_{-32.7}^{+46.1}$& $0.48_{-0.33}^{+0.39}$& $10.3_{-8.0}^{+11.8}$& \\
         C.1& $8.27_{-0.40}^{+0.43}$ & $18.2_{-12.5}^{+19.9}$& $0.08_{-0.06}^{+0.12}$& $2.2_{-1.4}^{+4.2}$& \\
         D.1& $8.66_{-0.62}^{+0.58}$ & $32.1_{-23.4}^{+49.8}$&  $0.31_{-0.23}^{+0.44}$& $4.0_{-3.1}^{+10.0}$& \\
         A.2& $8.71_{-0.36}^{+0.28}$ & $26.7_{-13.5}^{+14.5}$& $0.03_{-0.02}^{+0.05}$& $6.3_{-3.7}^{+6.1}$& \\
         B.2& $8.54_{-0.55}^{+0.50}$ & $27.9_{-19.4}^{+39.1}$& $0.30_{-0.21}^{+0.39}$& $3.5_{-2.6}^{+7.2}$& \\
         C.2& $8.63_{-0.53}^{+0.49}$ & $25.0_{-17.9}^{+33.7}$& $0.18_{-0.13}^{+0.23}$& $4.8_{-3.5}^{+10.1}$& \\
         \hline
    \end{tabular}
    \begin{tablenotes}  
    \footnotesize
    \item Notes. For the entire systems (ID1 and ID2), we use isophotal fluxes and fit only with \texttt{Prospector}. Stellar masses and SFRs are not corrected for magnification.
    \end{tablenotes}
    \label{tab:sed}
\end{table*}

\section{Results and Discussion}
\label{sec:res_diss}

\subsection{Physical Properties}
\label{subsec:property}

The entire systems of ID1 and ID2 have intrinsically faint UV luminosity $M_\mathrm{UV}=-17.9\pm0.2$, which corresponds to sub-$L_*$ \citep{Harikane2023}, placing them among the faintest galaxies known at $z>10$ (see Figure \ref{fig:z-M_UV}). The intrinsic stellar mass and half-light radius of ID1 (ID2) are $\log M_\mathrm{*,int}=3.7_{-1.7}^{+3.0}\times10^7\ M_\odot$ ($3.4_{-1.7}^{+2.2}\times10^7\ M_\odot$) and $\mathrm{r_{eff,int}}=181\pm18$ pc ($197\pm28$ pc), respectively. We estimate the half-light radii of ID1 and ID2 from the growth curve of their detection images (see Figure \ref{fig:plane}). We then obtain the intrinsic half-light radii by correcting for magnification with the Glafic model. We note that the choice of the models has minor effects on the stellar mass and half-light radius estimates, as demonstrated in Appendix \ref{sec:lens_uncertainty}. The sizes and UV luminosities of ID1 and ID2 are comparable to the size-UV luminosity relation for $z>10$ \citep{Ono2025}, which suggests that our galaxy is a representative example of faint, low-mass galaxies at $z>10$. As mentioned above, a remarkable feature of this galaxy is that due to the strong magnification, it is resolved into multiple clumps, the central one of which is more than two times brighter than the other clumps. At $z>10$, there exist a few studies about similar clumpy galaxies, BulletArc-z11 ($z_\mathrm{spec}=11.10$; \citealt{Bradac2025}) and MACS0647-JD ($z_\mathrm{spec}=10.17$; \citealt{Hsiao2023,Hsiao2024}), both of which show a bright central clump and the other smaller clumps. Interestingly, MACS0647-JD may be undergoing a galaxy merger based on the different SFHs among the clumps while the individual clumps of BulletArc-z11 exhibit similar SFHs. We further discuss the possible clump formation scenario in the next section.

In the left panel of Figure \ref{fig:size_density}, we present the size and stellar mass relation of local star clusters, star-forming clumps, and dwarf galaxies. The individual clumps of the Misty Moons are located in similar regions to local nuclear star clusters (NSCs), young massive star clusters (YMCs), ultra compact dwarfs (UCDs), and compact elliptical galaxies (cEs; \citealt{Norris2014}) as well as high-$z$ individual star-forming clumps of the BulletArc-z11 ($z_\mathrm{spec}=11.10$; \citealt{Bradac2025}) and Cosmic Grapes ($z_\mathrm{spec}=6.07$; \citealt{Fujimoto2025}). The Cosmic Gems arc ($z_\mathrm{spec}=9.63$; \citealt{Adamo2024,Bradac2024,Messa2026,Vanzella2026}) and Firefly Sparkle ($z_\mathrm{spec}=8.30$ \citealt{Mowla2024}) are further resolved into $<10$ pc scale star clusters, similar to local young star clusters (YSCs) and GCs. The studies of GCs in galaxy clusters at $z\sim0.3-0.9$ have become rather robust with JWST/NIRCam (see \citealt{Berkheimer2025,Harris2025}). The formation ages and masses of these low-$z$ GCs estimated from their colors are broadly consistent with those of our clumps, suggesting the possibility of being proto-GCs. We note that GCs are predicted to lose a large fraction of stars over time due to tidal friction in their host galaxies (e.g., \citealt{Harris2025}), likely explaining why local GCs have systematically lower stellar masses.
Combining the delensed stellar masses and SFRs with intrinsic half light radii, we derive the stellar mass and SFR surface densities. Some of the clumps show high surface densities of $\Sigma_*\sim10^{4}\ M_\odot\ \mathrm{pc^{-2}}$ and $\Sigma_\mathrm{SFR}\sim10^2\ M_\odot\ \mathrm{yr^{-1}}\ \mathrm{kpc^{-2}}$. \red{For the most compact clumps of A.1 and A.2, if they are unresolved, the surface densities may be even higher (see Section \ref{subsec:clump}).} These values are comparable to those of compact, nitrogen-rich galaxies at $z\gtrsim6$ (e.g., \citealt{Isobe2023b,Senchyna2024,Schaerer2024,Castellano2024}) although similarly high densities can also be found in a few non-nitrogen-rich galaxies \citep{Schaerer2024}.

As presented in Table \ref{tab:sed}, \texttt{Prospector} fitting results indicate that most of the clumps consist of young ($\sim40-70$ Myr) stars, formed in a low-dust ($A_\mathrm{V}\lesssim0.1$ mag) environment. This is consistent with the rising star formation history obtained for these clumps, which is likely more common in the high-redshift Universe \citep[e.g., ][]{Harikane2025,Tang2025}. The results from \texttt{Bagpipes} are broadly consistent with those from \texttt{Prospector}. The stellar ages for individual clumps estimated with non-parametric SFHs are relatively older than those estimated with short SFHs, although the differences are not conclusive given their uncertainties. If real, these differences may indicate that individual clumps contain older stellar populations. We cannot yet exclude the possibility that the central clumps (A.1 and A.2) host active galactic nuclei (AGNs), given their brightness (about half of the galaxy's total brightness) and compactness \red{while we note that they may be marginally resolved with respect to the PSF (see Section \ref{subsec:clump}).}
%We note that although the central bright, compact clumps of ID1 and ID2 (A.1 and A.2, respectively) might host active galactic nuclei (AGNs), they are marginally resolved from the PSF.

\subsection{Clump Formation}
\label{subsec:formation}

\begin{figure*}[ht!]
    \centering
    \includegraphics[width=0.99\linewidth]{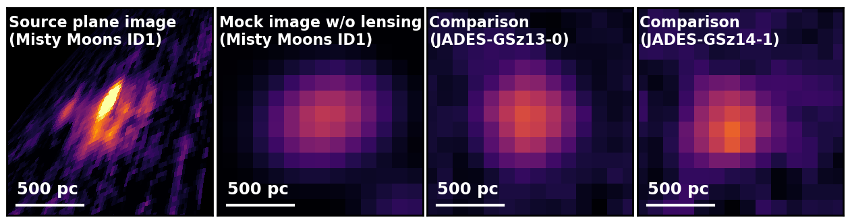}
    \caption{Comparison of the morphology of faint galaxies with and without lensing magnification. The leftmost panel presents the source plane image for ID1 of the Misty Moons. The second panel shows the mock image convolved with the NIRCam PSF. The third and rightmost figures represent the images of faint galaxies, GS-z13-0 ($z_\mathrm{spec}=13.20$; \citealt{Curtis-Lake2023}) and GS-z14-1 ($z_\mathrm{spec}=13.90$; \citealt{Carniani2024}). Without lensing magnification, ID1, GS-z13-0, and GS-z14-1 exhibit similar morphologies.}
    \label{fig:source_plane}
\end{figure*}

As shown in Figure \ref{fig:plane}, multiple clumps are resolved in the Misty Moons at this high redshift due to the high magnifications. To investigate the lensing effects, we construct a mock observation image of ID1 by convolving the source-plane image with the NIRCam PSF. The mock image shows smoothed morphology, yielding a light distribution comparable to that of similarly faint ($M_\mathrm{UV}\gtrsim-19$) galaxies at $z>10$, GS-z13-0 ($z_\mathrm{spec}=13.20$; \citealt{Curtis-Lake2023}) and GS-z14-1 ($z_\mathrm{spec}=13.90$; \citealt{Carniani2024}), as shown in Figure \ref{fig:source_plane}. \red{These resolution-dependent effects have also been reported in mock observations of galaxies from zoom-in cosmological simulations, where galaxy substructures identified at high spatial resolution become blended or undetectable when observed at lower resolution (e.g., \citealt{Zanella2021}).} As discussed for similarly strongly lensed galaxies at $z\gtrsim6$ \citep{Fujimoto2025,Vanzella2026}, this suggests that faint galaxies at $z>10$ likely have internal clumpy structure that remains unresolved without lensing magnifications.

\begin{figure*}
    \centering
    \includegraphics[width=0.99\linewidth]{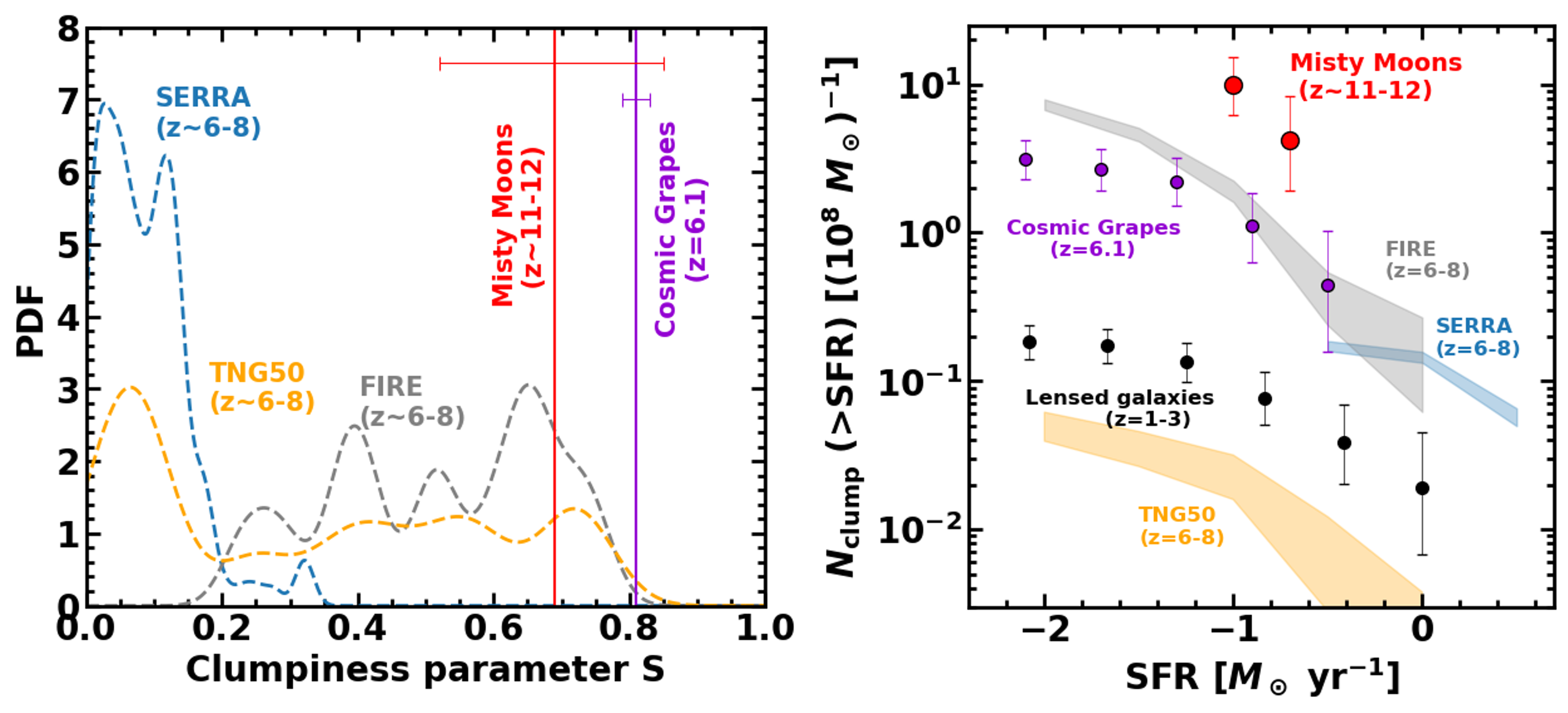}
    \caption{\red{Left: probability distributions of clumpiness parameters $S$. The red vertical and horizontal lines indicate the $S$ measurement and its $1\sigma$ uncertainty of the Misty Moons at $z\sim11-12$, respectively, while the purple lines show those for the Cosmic Grapes at $z=6.1$ \citep{Fujimoto2025}. The blue, yellow, and gray dashed curves show the distributions of the $S$ measurements for galaxies in the SERRA \citep{Pallottini2022}, TNG50 \citep{Pillepich2019}, and FIRE \citep{Wetzel2023} simulations at $z\sim6-8$, respectively, obtained by \citet{Fujimoto2025}. Right: cumulative clump luminosity function normalized by galaxy stellar mass as a function of clump star formation rate. The red, purple, and black circles present the measurements for the Misty Moons, Cosmic Grapes, and lensed galaxies at $z\sim1-3$ \citep{Livermore2015}, respectively. The blue, yellow, and gray shaded regions denote the measurements for the SERRA, TNG50, and FIRE galaxies, respectively.}}
    \label{fig:clumpiness}
\end{figure*}

\red{
To evaluate clumpiness in the Misty Moons, we derive the clumpiness parameter $S$ and clump luminosity function (LF), following \citet{Fujimoto2025}. The clumpiness parameter is defined as
\begin{equation}
    S=10\times\Sigma_{x,y=1,1}^{N,N}\frac{(I_{x,y}-I^\sigma_{x,y})-(B_{x,y}-B^\sigma_{x,y})}{I_{x,y}},
\end{equation}
where $I_{x,y}$ is the sky-subtracted flux values of the galaxy at position $(x,\ y)$, $I^\sigma_{x,y}$ is the flux values of the galaxy at position $(x,\ y)$ smoothed by Gaussian kernel with width $\sigma$, $B_{x,y}$ and $B^\sigma_{x,y}$ are the unsmoothed and smoothed flux values, respectively, but for background pixels in an area equal to that of the galaxy, and $N$ is the size of the galaxy in pixels \citep{Conselice2003}. To ensure sufficient S/N ratios in individual pixels, we use the detection image of ID1 (Section \ref{subsec:detection}) within a circular aperture of radius $r=0\farcs7$ centered on the peak pixel and adopt $\sigma=0\farcs2$, following \citet{Fujimoto2025}. We confirm that these values are well matched to the observed light distribution of ID1. The $1\sigma$ uncertainty of the $S$ value is derived from $1000$ Monte Carlo realizations of the observed image, where each pixel is perturbed according to its $1\sigma$ uncertainty assuming a Gaussian error distribution. We obtain $S=0.69_{-0.17}^{+0.16}$ for the Misty Moons, which is comparable to the high value measure for the Cosmic Grapes at $z=6.1$ \citep{Fujimoto2025} (left panel of Figure \ref{fig:clumpiness}). In the right panel, we present the cumulative clump LF as a function of SFR normalized by the galaxy stellar mass, allowing comparison among galaxies with different stellar masses. We calculate the number of clumps in the Misty Moons within each SFR bin using the SED-based SFRs, adopting the mean value of ID1 and ID2. We derive $1\sigma$ uncertainty based on the Poisson statistics (e.g., \citealt{Gehrels1986}). The clump LF of the Misty Moons is systematically higher than those of lensed galaxies at $z\sim1-3$ \citep{Livermore2015} and Cosmic Grapes ($z=6.1$). We further compare the clumpiness parameter and clump LF with those of simulated galaxies at $z\sim6-8$ analyzed by \citet{Fujimoto2025}, based on the cosmological zoom-in simulations of SERRA \citep{Pallottini2022}, TNG50\footnote{\url{https://www.tng-project.org/about/}} \citep{Pillepich2019}, and FIRE\footnote{\url{https://flathub.flatironinstitute.org/fire}} \citep{Wetzel2023}. As described in \citet{Fujimoto2025}, the selected SERRA, TNG50, and FIRE galaxies show gas kinematics with likely rotating disk, asymmetric velocity gradient, and disturbed gas without major mergers, respectively. The Misty Moons exhibit higher clumpiness than these simulated galaxies in both the $S$ values and clump LFs, similar to the Cosmic Grapes. For the Cosmic Grapes, one plausible explanation for the numerous clumps despite the presence of a rotating disk is substantially weak feedback \citep{Fujimoto2025}. Although whether a rotating disk exists in the Misty Moons cannot be confirmed with the current data, the weak feedback scenario may still be applicable to the Misty Moons as well, given its high clumpiness comparable to that of the Cosmic Grapes. The comparisons of the clumpiness among the observed and simulated galaxies across cosmic time suggest that the galaxies at higher redshift tend to be more clumpy, which is consistent with the observations of many clumpy galaxies at $z>6$ (e.g., \citealt{Adamo2024,Bradley2024,Mowla2024,Bradac2025,Fujimoto2025,Harikane2025,Messa2025,Messa2026,Vanzella2026}). The elevated clumpiness is qualitatively consistent with a large fraction of star formation being concentrated in compact clumps, motivating a direct examination of the clustered stellar mass fraction.
}

%In the left panel of Figure \ref{fig:clump_formation}, we compare clump mass fractions of high-$z$ clumpy galaxies (\citealt{Vanzella2022,Adamo2024,Bradley2024,Bradac2025,Fujimoto2025,Messa2026,Vanzella2026}; This work) with cluster mass fractions of local dwarf galaxies \citep{Calzetti2015,Cook2023}. We derive the clump mass fraction, dividing the sum of the clump masses by total galaxy mass. For cluster mass fraction, we first calculate cluster masses by multiplying cluster formation rate of CFR ($1-100$ Myr) in Table 4 of \citet{Cook2023}  by $100$Myr. We then obtain the cluster mass fractions, dividing the cluster masses by total masses presented in \citet{Calzetti2015}. The high-$z$ clumpy galaxies have approximately two to three order higher mass fractions ($\gtrsim30\%$) than mass fractions of local star clusters ($\lesssim1\%$). This high mass fraction at high-$z$ is more preferred in the lower-mass galaxies, as shown in the increasing mass fraction for lower stellar mass (brown star symbols in the left panel of Figure \ref{fig:clump_formation}; \citealt{Claeyssens2025}), which is also theoretically suggested \citep{Elmegreen2018}. This indicates that star-forming clumps and star clusters at high-$z$ dominate the mass budget of the entire systems. In other words, high-$z$ low-mass clumpy galaxies might be in a totally different clump/cluster formation mode (top down star formation), in contrast to the local star cluster \citep[e.g., ][]{Cook2023}.
\begin{figure*}
    \centering
    \includegraphics[width=0.99\linewidth]{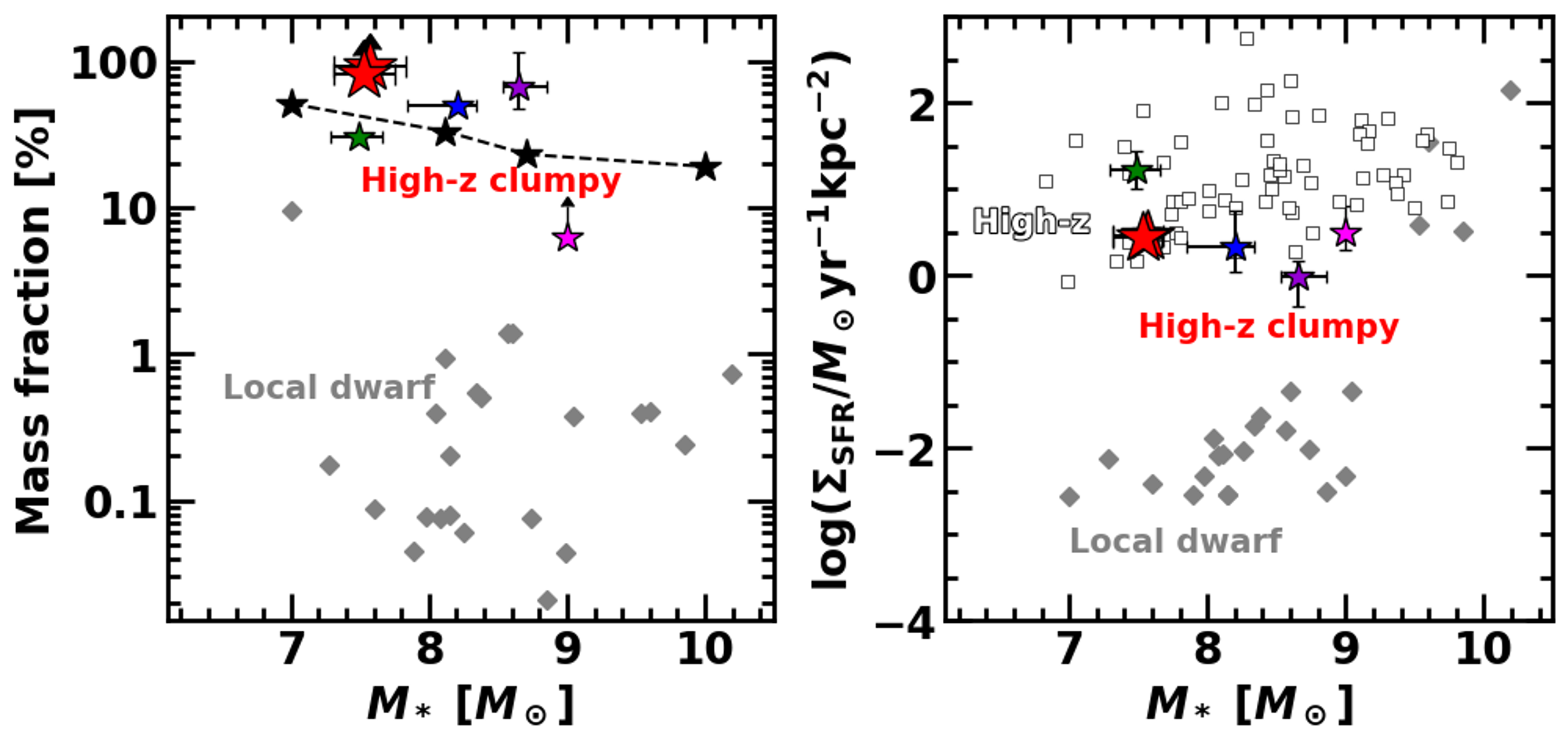}
    \caption{Left: mass fraction in clustered stellar systems as a function of stellar mass. The red star symbols indicate ID1 and ID2 of the Misty Moons ($z_\mathrm{phot}\sim11-12$). The blue, green, purple, and magenta star symbols show the BulletArc-z11 \citep[$z_\mathrm{spec}=11.10$;][]{Bradac2025}, Cosmic Gems arc \citep[$z_\mathrm{spec}=9.63$;][]{Adamo2024,Bradley2024,Messa2026,Vanzella2026}, Cosmic Grapes \citep[$z_\mathrm{spec}=6.07$;][]{Fujimoto2025}, and Sunburst galaxy \citep[$z_\mathrm{spec}=2.37$;][]{Vanzella2022}, respectively. The black star symbols denote the median values for clumpy galaxies at $1<z<5.5$ \citep{Claeyssens2025}. The gray diamonds present the local dwarf galaxies containing young star clusters \citep{Adamo2011,Calzetti2015,Cook2023}. Right: SFR surface density as a function of stellar mass. The star symbols and diamonds are the same in the left panel. The squares represent spectroscopically confirmed galaxies at $5<z<14$ \citep{Morishita2024}.
    %Cluster formation efficiency as a function of SFR surface density (left) and stellar mass (right). The red star symbols indicate ID1 and ID2 of MJ0257-2325-z12. The blue, green, purple, and magenta circles show the BulletArc-z11 \citep[$z=11.10$;][]{Bradac2025}, Cosmic Gems arc \citep[$z=9.63$;][]{Adamo2024,Bradley2024,Messa2025,Vanzella2026}, Cosmic Grapes \citep[$z=6.07$;][]{Fujimoto2025}, and Sunburst galaxy \citep[$z=2.37$;][]{Vanzella2022}, respectively. The black circles and squares present the measurements of the local galaxies and their median values, respectively, compiled by \citet{Adamo2020}. The red and blue square denote the simulation results for low-mass ($M_*\sim10^{7-9}\ M_\odot$) and high-mass ($M_*\sim10^{9-11}\ M_\odot$) galaxies in the local Universe, respectively \citep{Pfeffer2019}, while the black dashed line represent the analytical model of \citet{Kruijssen2012}. 
    }
    \label{fig:clump_formation}
\end{figure*}

In the left panel of Figure \ref{fig:clump_formation}, we compare the clustered  mass fractions of high-$z$ galaxies (colored symbols; \citealt{Vanzella2022,Adamo2024,Bradley2024,Bradac2025,Claeyssens2025,Fujimoto2025,Messa2026,Vanzella2026}; this work) with cluster mass fractions of local galaxies (gray symbols; blue compact dwarfs (BCDs): \citealt{Ostlin2001,Adamo2011} and low-mass dwarfs: \citealt{Calzetti2015,Cook2023}). We derive the clustered mass fraction, dividing the sum of the clump/cluster masses by total galaxy mass. For the Misty Moons, the probability distributions of the mass fractions pile up near unity. We thus obtain $1\sigma$ lower limits of $90\%$ for ID1 and $78\%$ for ID2, using the total stellar masses derived from the total isophotal fluxes. While the isophotal regions include light from a diffuse component that may contain older stellar populations, this contribution could be overwhelmed by the light from young star-forming clumps, which is known as stellar outshining (e.g., \citealt{Narayanan2024}). Since the stellar outshining may lead to an underestimate of the total stellar mass, we also derive the clump mass fractions of the Misty Moons, using diffuse fluxes explicitly estimated from the \texttt{Galfit} fitting (Section \ref{subsec:clump}). The resulting clump mass fractions are $54_{-24}^{+42}$\% for ID1 and $44_{-23}^{+38}$\% for ID2. The differences of the estimates from those derived with the total isophotal fluxes are within factor of two, which does not change our conclusion based on differences by orders of magnitude. In the following discussion, we adopt the mass fractions based on the total isophotal fluxes as fiducial results of the Misty Moons because the stellar masses estimated from the diffuse fluxes have larger uncertainties due to their low surface brightnesses. For the cluster mass fractions of the Cook et al. sample, we first calculate cluster masses by multiplying cluster formation rate of CFR ($1-100$ Myr) in Table 4 of \citet{Cook2023}  by $100$ Myr. We then obtain the cluster mass fractions, dividing the cluster masses by total masses presented in \citet{Calzetti2015}. Similarly, we derive the mass fraction of luminous BCDs (Haro11 and ESO338) from \citet{Adamo2011}, using galaxy masses from \citet{Ostlin2001}. In the gathered sample at all redshifts, we effectively compare the recently formed mass (within few hundreds Myr) in clustered (star clusters/clump) systems to the total mass of the galaxy. The high-$z$ galaxies have approximately two to three dex higher mass fractions ($\gtrsim30\%$) than that measured in  local galaxies ($\lesssim1\%$). Such high mass fractions are emerging as a preferred trait in lower-mass galaxies at high-$z$ \citep{Adamo2025}, as shown by the increasing mass fraction for lower stellar mass (black star symbols in the left panel of Figure \ref{fig:clump_formation}; \citealt{Claeyssens2025}). % which is also theoretically suggested \citep{Elmegreen2018}. 
Occasionally, due to accretions or merger events, local dwarf galaxies experience an increase in gas pressure and SFR density, resulting in the formation of massive star clusters. However, these events take place in a system where the stellar mass is typically dominated by stellar populations formed over a long timescale. In contrast, for high-$z$ low-mass galaxies, star-forming clumps and star clusters dominate the stellar mass budget of the entire systems. In other words, high-$z$ low-mass clumpy galaxies might be in a totally different star formation mode, dominated by top down clustered star formation \citep{Elmegreen2018}.%, in contrast to local galaxies where the current star formation proceeds at slower pace and star cluster formation  \citep[e.g., ][]{Cook2023}.

One of the key differences between the high-$z$ clumpy galaxies and local dwarf galaxies is SFR surface density. As shown in the right panel of Figure \ref{fig:clump_formation}, the high-$z$ clumpy galaxies exhibit $\sim10^2-10^4$ times higher SFR surface densities than those of the local dwarf galaxies with similarly low stellar masses of $M_*\sim10^7-10^9M_\odot$. Considering the positive correlation between cluster formation efficiency and SFR surface density in the local galaxies reported by both observational (e.g., \citealt{Adamo2020}) and theoretical (e.g., \citealt{Kruijssen2012,Pfeffer2019}) studies, the high SFR density may be a \red{possible} driver of the high mass fraction seen in the high-$z$ clumpy galaxies. However, we note that the local BCDs with higher stellar masses of $M_*>10^9M_\odot$ have high SFR densities but low mass fractions compared to the high-$z$ clumpy galaxies. This suggests that high-$z$ clumpy galaxies are likely distinct from the local dwarf galaxies not only in their high SFR surface densities but also in their short, bursty star formation during the early formation phase of the galaxy. 

High SFR surface density is associated with high gas density (e.g., \citealt{Kennicutt2012}), which is a major driver of the violent disk instability, one possible scenario for clump formation at high-$z$ (e.g., \citealt{Romeo2010,Dekel2009,Dekel2013,Romeo2014,Fujimoto2025}). Recent simulation results of \citet{Mayer2025} have shown that low-mass ($\sim10^8\ M_\odot$) galaxies form massive self-gravitating compact gas disks, which collapse into massive bound clumps by gravitational instability. We note that although it seems difficult to form stellar disks at this high redshift, some studies suggest possible rotating gas disk formation at $z>10$ \citep[e.g., ][]{Xu2024,Scholtz2025}. The disk instability scenario is consistent with the top-down clump formation suggested by high clustered mass fractions. Regarding the in-situ clumps formation, filament fragmentation (e.g., \citealt{Garcia2025}) and turbulence-driven fragmentation (e.g., \citealt{Dessauges-Zavadsky2018}) are also suggested as possible scenarios. In recent studies using a cosmological radiation-hydrodynamic simulation \citep{Garcia2023,Garcia2025}, low mass dwarf galaxies at $z>8$ with no disk and bulge experience star cluster formation through the fragmentation of accreting dense filaments (i.e., cold inflow) before the gas reached the center. The suppression of cooling via the far-UV (FUV) radiation leads to bursty star formation and massive, compact star clusters \citep{Sugimura2024}, sharing the comparable properties with high-$z$ star clusters \citep{Mowla2024,Vanzella2026}. These star clusters orbit the center of the dark matter halo, where star formation happens later, resulting in the formation of a NSC \citep{Garcia2025}. Considering that our central clumps (A.1 and A.2) have comparable stellar masses and effective radii to those of NSCs (Figure \ref{fig:size_density}), the filament fragmentation is also a plausible scenario for the clump formation of the Misty Moons.

One of the other possibilities of clump formation is a galaxy merger (e.g., \citealt{Matthee2017,Hsiao2023,Boyett2024,Marconcini2024,Nakazato2024,Harikane2025,Kiyota2025,Bik2026,Pascalau2026}). In fact, \citet{Hsiao2023} suggest that the observed clumpy structure of MACS0647-JD at $z_\mathrm{spec}=10.17$ could be due to a galaxy merger based on the different SFHs among the clumps and comparison with some simulation results. In a recent study of \citet{Kohandel2025}, they identify Amaryllis, a digital twin of $z>10$ galaxies, in the SERRA zoom-in simulations. At $z>10$, Amaryllis has a main progenitor and coexisting small companions, which form stars in a clustered and bursty way, and
eventually merge. The morphology of Amaryllis is very similar to the Misty Moons (see Figure 1 in \citealt{Kohandel2025}). Observationally, the Misty Moons has stellar ages for the central clumps (A.1 and A.2) are relatively younger than the other clumps (see Table \ref{tab:sed}), which might suggest a possible merger scenario. However, the current NIRCam images have almost no coverage of the wavelengths longer than Balmer breaks, which complicates the exact measurements of stellar ages and SFHs enough to conclude that the individual clumps undergo different SFHs. Future observations with NIRSpec/IFU (for gas kinematics and spatial distribution of stellar/gas properties) and MIRI (for metallicity measurement and optical continuum from old stellar populations) are helpful to further explore the clump properties and physical origins of the clump formation.

\section{Summary} 
\label{sec:summary}
In this paper, we report the discovery of the \red{Misty Moons, a strongly lensed galaxy at $z\sim11-12$ behind the galaxy cluster MACS J0257.1-2325}, newly identified in the JWST Treasury Survey VENUS. Independent lens models predict that the Misty Moons has five multiple images, two of which, ID1 and ID2 ($\mu\sim20-30$), are very bright ($\mathrm{F200W}\sim26$ mag) in the image plane, while intrinsically faint ($M_\mathrm{UV}\sim-18$) in the source plane, and exhibit blue SEDs with prominent Ly$\alpha$ breaks. In addition, ID1 and ID2 are resolved into multiple clumps in NIRCam images. We summarize our main findings below:

\begin{itemize}
    \item[1.] 
    We conduct SED fitting to ID1 and ID2 with \texttt{EAZY} and \texttt{Prospector} to estimate their photometric redshifts and stellar populations. For both the fitting codes, we adopt models excluding and including Ly$\alpha$ emission, resulting in $z_\mathrm{phot}\sim11$ and $z_\mathrm{phot}\sim12$, respectively. We also fit the individual clumps of ID1 and ID2, fixing the redshift to $z=11.5$. The SED fitting results show that the entire systems and individual clumps of the Misty Moons consist of young stars ($t\sim40-70$ Myr with non-parametric SFHs and $t\sim20-40$ Myr with short SFHs) with low dust content ($A_\mathrm{V}\lesssim0.1$ mag), likely undergoing their first significant starbursts.
    
    \item[2.] 
    Correcting for the high magnification, ID1 (ID2) is resolved into four (three) clumps with effective radii of $r_\mathrm{eff}\sim10-70$ pc and stellar masses of $M_*\sim10^7\ M_\odot$, which are comparable to those of the local nuclear/young massive star clusters and high-$z$ ($z\gtrsim6$) star-forming clumps. \red{The Misty Moons exhibit a high clumpiness parameter and clump LF normalized by galaxy stellar mass compared to lower-redshift lensed galaxies and simulated galaxies.} In particular, these star-cluster-like clumps have high clustered mass fractions ($\gtrsim80\%$), similar to other high-$z$ clumps and star clusters. This suggests that star formation \red{in at least some low-mass galaxies at these early epochs may be strongly} clump-dominated, in contrast to compact star clusters observed in local dwarf galaxies, which might be driven by high SFR surface densities and short star formation timescales.
    
    \item[3.]
    We convolve the source-plane image with the JWST/NIRCam point-spread function to produce a mock NIRCam image of the Misty Moons without lensing magnification, and find that the intrinsic galaxy has a radial surface-brightness profile comparable to those of typical faint ($M_\mathrm{UV}\gtrsim-19$ mag) galaxies at $z\gtrsim10$, including JADES-GS-z13-0 and JADES-GS-z14-1. This suggests \red{that clump formation may be more common among} the low-mass galaxies in the early Universe.
\end{itemize}

%% IMPORTANT! The old "\acknowledgment" command has be depreciated. It was
%% not robust enough to handle our new dual anonymous review requirements and
%% thus been replaced with the acknowledgment environment. If you try to 
%% compile with \acknowledgment you will get an error print to the screen
%% and in the compiled pdf.
%% 
%% Also note that the akcnowlodgment environment does not support long amounts of text. If you have a lot of people and institutions to acknowledge, do not use this command. Instead, create a new \section{Acknowledgments}.

\section*{Acknowledgements} 
We thank anonymous referee for the valuable comments that improved this manuscript. We are grateful to the JADES, RELICS, and VENUS teams for developing their imaging surveys, enabling our research. \red{We thank Mahsa Kohandel and Andrea Pallottini for providing SERRA simulation data and the useful discussion on stellar clumps.} We thank Yurina Nakazato for the valuable discussions on clump formation. We thank Bingjie Wang for the helpful discussions on \texttt{Prospector} SED fitting. This work is based on observations made with the NASA/ESA/CSA James Webb Space Telescope. The data were obtained from the Mikulski Archive for Space Telescopes (MAST) at the Space Telescope Science Institute (STScI), which is operated by the Association of Universities for Research in Astronomy (AURA), Inc., under NASA contract NAS 5-03127 for JWST. The JWST observations are associated with programs \#6882 and \#1180 (JADES). This paper is also based on observations made with the NASA/ESA Hubble Space Telescope. The data were obtained from MAST at STScI, which is operated by AURA under NASA contract NAS 5-26555. The HST observations are associated with programs \#14096 (RELICS). The authors acknowledge the use of the Canadian Advanced Network For Astronomy Research (CANFAR) Science Platform operated by the Canadian Astronomy Data Center (CADC) and the Digital Research Alliance of Canada (DRAC) with support from the National Research Council for Canada (NRC), the Canadian Space Agency (CSA), CANARIE, and the Canadian Foundation for Innovation(CFI). 
We acknowledge the support of the Canadian Space Agency (CSA) [25JWGO4A18]. 
MN acknowledges support from  KAKENHI Grant Nos. 25KJ0828 through Japan Society for the Promotion of Science (JSPS).
SF acknowledges support from the Dunlap Institute, funded through an endowment established by the David Dunlap family and the University of Toronto.
MO acknowledges the supports from the World Premier International Research Center Initiative (WPI Initiative), MEXT, Japan, the joint research program of the Institute for Cosmic Ray Research (ICRR), the University of Tokyo, and KAKENHI (20H00180, 21H04467, 25H00674) through JSPS.
DM acknowledges the support of the Royal Society through the award of a Royal Society University Research Professorship to James S. Dunlop.
MO acknowledges the supports from JSPS KAKENHI Grant Numbers JP25H00662, JP22K21349.
AZ acknowledges support by the Israel Science Foundation Grant No. 864/23.
AA acknowledges support by the Swedish research council Vetenskapsr{\aa}det (VR) project 2021-05559, and VR consolidator grant 2024-02061.
EV and MM acknowledge financial support through grants PRIN-MIUR 2020SKSTHZ, the INAF GO Grant 2024 ``Mapping Star Cluster Feedback in a Galaxy 450 Myr after the Big Bang'', and by the European Union – NextGenerationEU within PRIN 2022 project n.20229YBSAN - Globular clusters in cosmological simulations and lensed fields: from their birth to the present epoch.
MB and GR acknowledge support from the ERC 848 Grant FIRSTLIGHT and Slovenian national research 849 agency ARIS through grants N1-0238 and P1-0188.
MM acknowledges financial support through grants INAF GO Grant 2022 ``The revolution is around the corner: JWST will probe globular cluster precursors and Population III stellar clusters at cosmic dawn.'' 
HY acknowledges support by KAKENHI (25KJ0832) through JSPS.
RA acknowledges support of Grant PID2023-147386NB-I00 funded by MICIU/AEI/10.13039/501100011033 and by ERDF/EU, and  the Severo Ochoa award to the IAA-CSIC CEX2021-001131-S and from grant PID2022- 136598NB-C32 "Estallidos8."
HA acknowledges support from CNES, focused on the JWST mission, and the French National Research Agency (ANR) under grant ANR-21-CE31-0838.
CJC acknowledges support from the ERC Advanced Investigator Grant EPOCHS (788113).
PD warmly acknowledges support from an NSERC discovery grant (RGPIN-2025-06182).
YH acknowledges support from the JSPS Grant-in-Aid for Scientific Research (24H00245) and the JSPS International Leading Research (22K21349).
YF acknowledges support from JSPS KAKENHI Grant Numbers JP22K21349 and JP23K13149.
YJ-T acknowledges financial support from the State Agency for Research of the Spanish MCIU through Center of Excellence Severo Ochoa award to the Instituto de Astrof\'isica de Andaluc\'ia CEX2021-001131-S funded by MCIN/AEI/10.13039/501100011033, and from the grant PID2022-136598NB-C32 Estallidos and project ref. AST22-00001-Subp-15 funded by the EU-NextGenerationEU.
GEM acknowledges the Villum Fonden research grants 13160 and 37440 and the Cosmic Dawn Center of Excellence funded by the Danish National Research Foundation under the grant No. 140.
GN is supported by the Canadian Space Agency under a contract with NRC Herzberg Astronomy and Astrophysics.
This work has received funding from the Swiss State Secretariat for Education, Research and Innovation (SERI) under contract number MB22.00072, as well as from the Swiss National Science Foundation (SNSF) through project grant 200020\_207349. The Cosmic Dawn Center (DAWN) is funded by the Danish National Research Foundation under grant DNRF140.
RAW acknowledges support from NASA JWST Interdisciplinary Scientist grants NAG5-12460, NNX14AN10G and 80NSSC18K0200 from GSFC.
FEB acknowledges support from ANID-Chile BASAL CATA FB210003, FONDECYT Regular 1241005, ECOS-ANID ECOS240037, and Millennium Science Initiative, AIM23-0001.
KK acknowledges support from JSPS KAKENHI Grant Numbers JP22H04939, JP23K20035, and JP24H00004.  
RPN is grateful for the generous support of Neil and Jane Pappalardo through the MIT Pappalardo Fellowship.
\red{The English in this paper was partially refined with the assistance of ChatGPT (OpenAI).}

%% To help institutions obtain information on the effectiveness of their 
%% telescopes the AAS Journals has created a group of keywords for telescope 
%% facilities.
%
%% Following the acknowledgments section, use the following syntax and the
%% \facility{} or \facilities{} macros to list the keywords of facilities used 
%% in the research for the paper.  Each keyword is check against the master 
%% list during copy editing.  Individual instruments can be provided in 
%% parentheses, after the keyword, but they are not verified.

%% Similar to \facility{}, there is the optional \software command to allow 
%% authors a place to specify which programs were used during the creation of 
%% the manuscript. Authors should list each code and include either a
%% citation or url to the code inside ()s when available

\software{NumPy \citep{Harris2020}, matplotlib \citep{Hunter2007}, SciPy \citep{Virtanen2020}, Astropy \citep{Astropy2013,Astropy2018,Astropy2022}, \texttt{grizli} \citep{Brammer2021,Brammer2023}, \texttt{Photutils} \citep{Bradley2025}, \texttt{EAZY} \citep{Brammer2008}, \texttt{Prospector} \citep{Johnson2021}, Glafic \citep{Oguri2010,Oguri2021}, \texttt{Bagpipes}, \citep{Carnall2018}, \texttt{GALFIT} \citep{Peng2010}}
%, emcee \citep{Foreman2013}}.

%% Appendix material should be preceded with a single \appendix command.

%% There should be a \section command for each appendix. Mark appendix
%% subsections with the same markup you use in the main body of the paper.

%% Each Appendix (indicated with \section) will be lettered A, B, C, etc.
%% The equation counter will reset when it encounters the \appendix
%% command and will number appendix equations (A1), (A2), etc. The
%% Figure and Table counter will not reset.

%% For this sample we use BibTeX plus aasjournals.bst to generate the
%% the bibliography. The sample631.bib file was populated from ADS. To
%% get the citations to show in the compiled file do the following:
%%
%% pdflatex sample631.tex
%% bibtext sample631
%% pdflatex sample631.tex
%% pdflatex sample631.tex

%\clearpage
%\bibliographystyle{aasjournal}
%\bibliographystyle{aasjournalv7}
\bibliography{library.bib}

%% This command is needed to show the entire author+affiliation list when
%% the collaboration and author truncation commands are used.  It has to
%% go at the end of the manuscript.
%\allauthors

%% Include this line if you are using the \added, \replaced, \deleted
%% commands to see a summary list of all changes at the end of the article.
%\listofchanges

% \clearpage
\appendix
\restartappendixnumbering
\section{Photometry}

In Table \ref{tab:photometry}, we list the photometry of the entire systems and individual clumps for ID1 and ID2 of the Misty Moons. The method is described in Section \ref{subsec:photometry}.

\section{Surface Brightness Profiles of the Most Compact Clumps}

\begin{figure*}[ht!]
    \centering
    \includegraphics[width=0.99\linewidth]{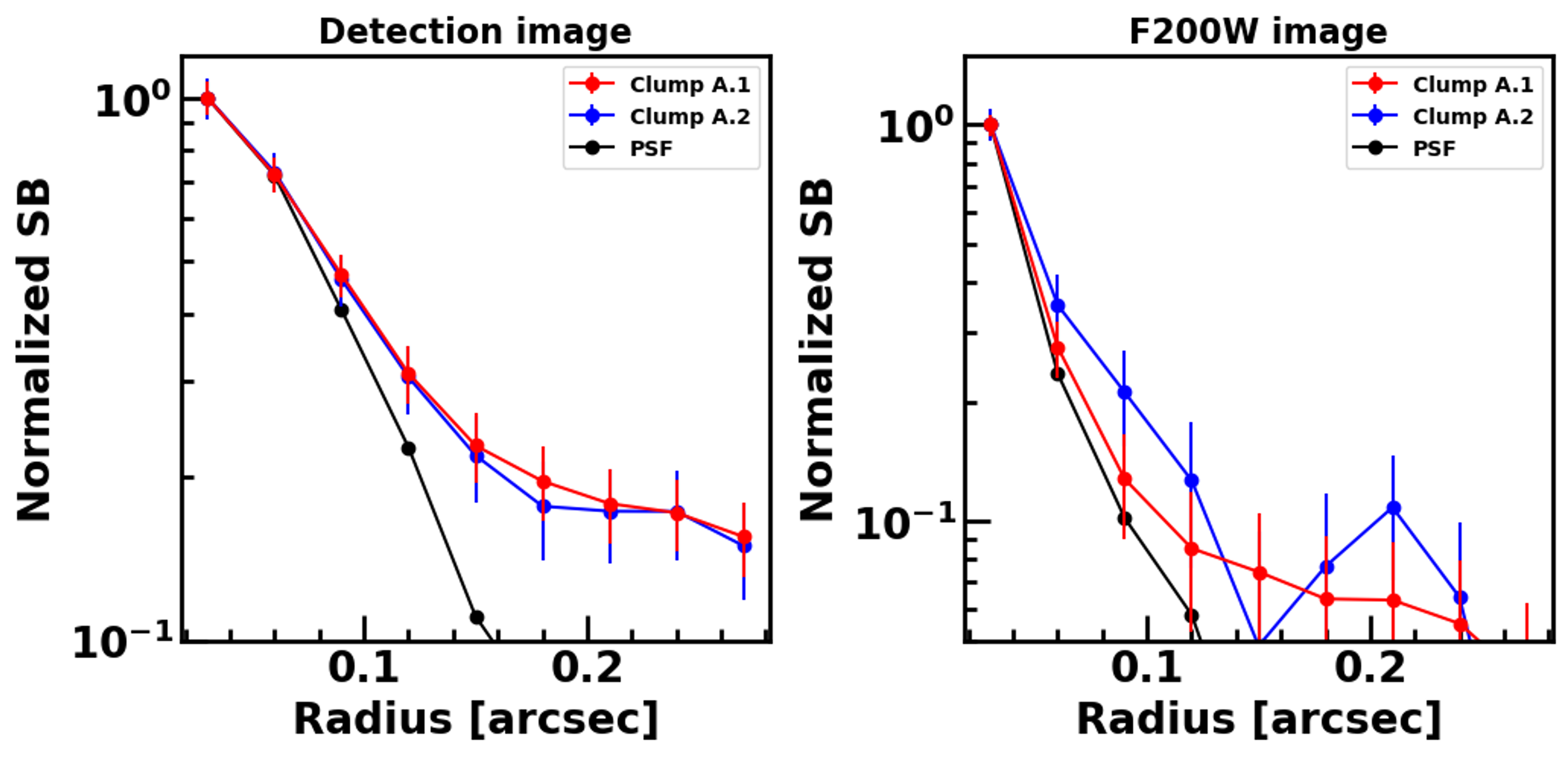}
    \caption{\red{Radial surface-brightness profiles of the most compact clumps A.1 and A.2 in the detection (F200W + F277W + F356W + F444W; left) and F200W images (right). The red, blue, and black circles present the measurements for clumps A.1, A.2, and PSF, respectively.}}
    \label{fig:sb_profile}
\end{figure*}

\red{In Section \ref{subsec:clump}, we investigate whether the most compact clumps of A.1 and A.2 are spatially resolved by using both S\'ersic-profile and pure-PSF fitting to the detection (F200W + F277W + F356W + F444W) and F200W images. To visually examine the light distribution, we measure radial surface brightness profiles of PSF, clumps A.1, and A.2 by measuring surface brightness in circular annuli centered on the brightest pixels using \texttt{Photutils}. The uncertainties are estimated from the corresponding variance images. The resulting surface-brightness profile are presented in Figure \ref{fig:sb_profile}. As described in Section \ref{subsec:clump}, the clumps are marginally broader than the PSF in the detection image while their profiles in the F200W image are consistent with the PSF within the measurement uncertainties.}

\label{sec:sb_profile}

\section{Effect of Lens Model Uncertainties on Derived Physical Properties}
\label{sec:lens_uncertainty}

\begin{figure*}[ht!]
\centering
    \includegraphics[width=0.99\linewidth]{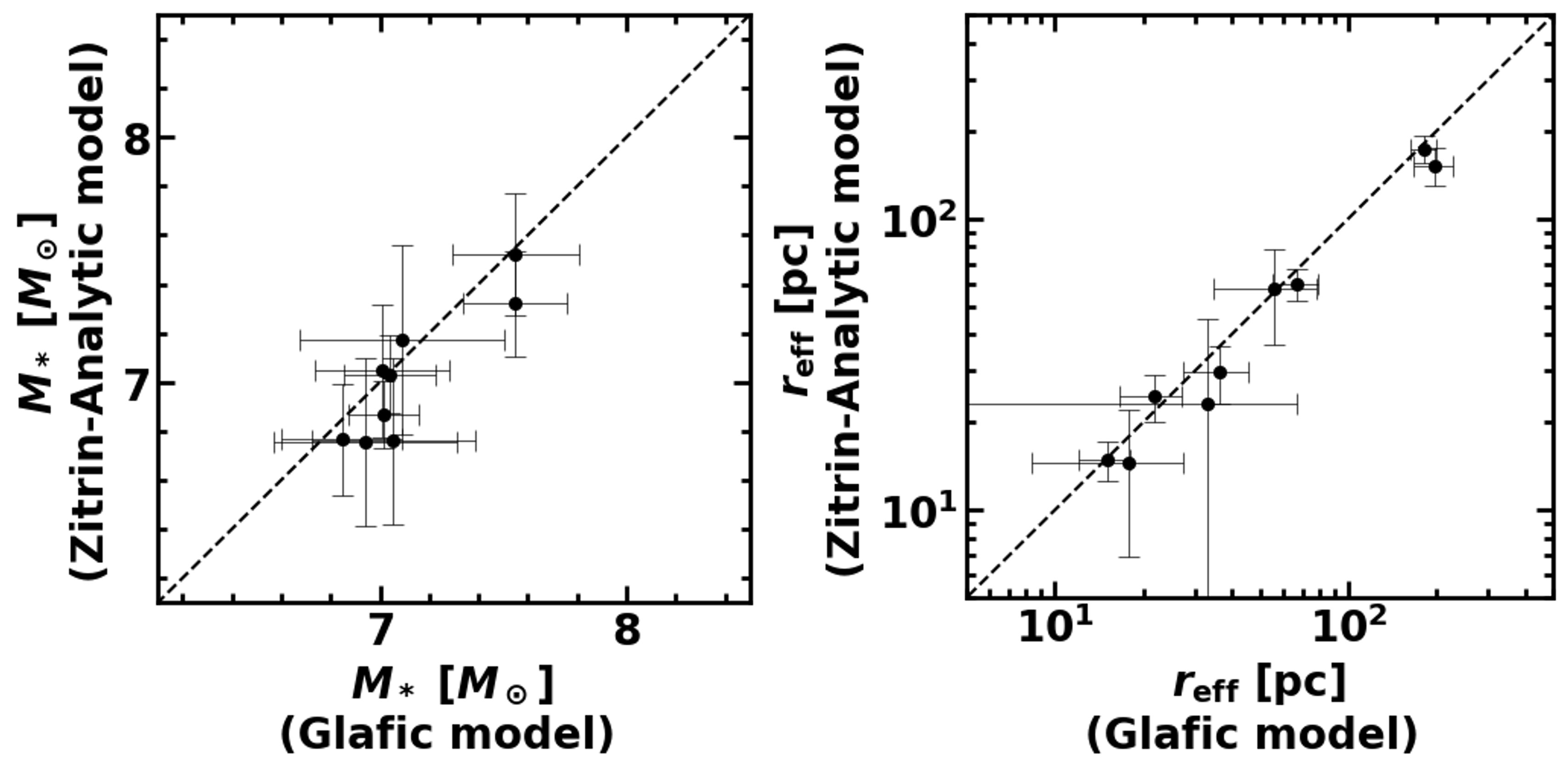}
    \caption{Comparison of intrinsic stellar mass and half-light radius measurements with Zitrin-Analytic and Glafic lens models. The circles present the measurements for the stellar mass (left) and half-light radius (right), and their $1\sigma$ uncertainties. The black lines indicate that the physical properties obtained from the two lens models are consistent.}
    \label{fig:lens_comparison}
\end{figure*}

This work is based on two independent lens models described in Section \ref{sec:lensmodel}. To evaluate the impact of lens model uncertainties on the intrinsic physical properties, we compare the delensed stellar masses and half-light radii derived with the Zitrin-Analytic and Glafic models in Figure \ref{fig:lens_comparison}. The figure shows that the measurements based on the two lens models are consistent within the uncertainties, indicating that the choice of lens model has only a minor impact on our discussion. We note that the surface density estimates are largely insensitive to the lens models because gravitational lensing conserves surface brightness to first order.

\setcounter{table}{0}
\renewcommand{\thetable}{A\arabic{table}}
\begin{longrotatetable}
\begin{table*}[h]
    \scriptsize
    \begin{center}
    \caption{Measured photomery of the Misty Moons.}
    \makebox[\textwidth][c]{
    \hspace{0.8cm}
    \begin{tabular}{lccccccccccccc}
    \hline
    \hline
    Name&F435W&F555W&F814W&F090W&F115W&F150W&F200W&F210M&F277W&F300M&F356W&F410M&F444W\\
    &(nJy)&(nJy)&(nJy)&(nJy)&(nJy)&(nJy)&(nJy)&(nJy)&(nJy)&(nJy)&(nJy)&(nJy)&(nJy)\\
    \hline
    %ID1 (Isophoto) & $-17.2\pm24.8$ & $9.2\pm10.5$ & $10.8\pm8.6$ & $4.5\pm12.6$ & $-4.5\pm13.3$ & $75.0\pm14.8$ & $134.2\pm10.3$ & $140.1\pm12.3$ & $125.0\pm6.4$ & $117.2\pm8.3$ & $110.3\pm6.1$ & $80.1\pm8.4$ & $109.6\pm8.4$ \\
    ID1 (Customized) & $-35.3\pm25.4$ & $5.3\pm9.1$ & $9.7\pm7.9$ & $2.9\pm9.5$ & $-3.7\pm10.7$ & $80.9\pm13.7$ & $129.8\pm13.4$ & $137.4\pm15.3$ & $118.4\pm10.5$ & $114.0\pm12.4$ & $104.7\pm9.4$ & $77.6\pm10.5$ & $106.0\pm10.6$ \\
    ID1 (Isophoto) & $-35.9\pm33.6$ & $-0.2\pm11.9$ & $19.1\pm12.0$ & $8.6\pm12.3$ & $2.9\pm13.3$ & $51.6\pm14.3$ & $133.0\pm10.6$ & $113.1\pm14.4$ & $118.4\pm6.8$ & $106.9\pm8.2$ & $110.9\pm6.0$ & $65.6\pm9.6$ & $125.5\pm6.8$ \\
    ID1 (Kron) & $-16.6\pm25.2$ & $12.5\pm13.7$ & $12.0\pm8.9$ & $3.5\pm9.1$ & $-3.3\pm10.7$ & $82.3\pm11.4$ & $151.7\pm9.9$ & $156.7\pm11.5$ & $130.1\pm8.7$ & $116.6\pm9.1$ & $109.5\pm7.6$ & $74.6\pm9.8$ & $96.2\pm8.3$ \\
    A.1 & - & - & - & - & - & $45.7\pm8.0$ & $63.1\pm5.2$ & $72.4\pm5.3$ & $53.5\pm4.9$ & $55.5\pm5.6$ & $47.4\pm4.8$ & $45.7\pm6.3$ & $47.4\pm4.8$ \\
    B.1 & - & - & - & - & - & - & $20.0\pm5.0$ & $23.3\pm5.6$ & $21.3\pm5.8$ & $24.4\pm5.9$ & $23.1\pm6.0$ & - & $28.6\pm5.0$ \\ 
    C.1 & - & - & - & - & - & $32.5\pm12.2$ & $48.3\pm7.6$ & $31.3\pm6.7$ & $18.2\pm7.9$ & $15.4\pm6.9$ & $25.8\pm8.0$ & - & $14.2\pm6.9$ \\ 
    D.1 & - & - & - & - & - & - & $24.4\pm7.1$ & - & $15.8\pm5.4$ & $22.9\pm7.7$ & $14.2\pm5.3$ & - & $17.5\pm7.1$ \\ 
    %ID2 (Isophoto) & $-19.4\pm25.4$ & $-0.7\pm10.6$ & $13.7\pm8.2$ & $11.2\pm12.1$ & $24.2\pm12.1$ & $70.4\pm14.1$ & $154.9\pm10.0$ & $114.1\pm12.5$ & $117.1\pm6.6$ & $111.2\pm8.7$ & $96.2\pm6.2$ & $102.9\pm9.0$ & $90.9\pm9.0$ \\
    ID2 (Customized) & $-17.3\pm26.0$ & $-2.8\pm8.7$ & $11.0\pm8.5$ & $10.7\pm10.5$ & $21.8\pm10.8$ & $69.3\pm13.0$ & $146.3\pm14.8$ & $110.8\pm14.7$ & $110.7\pm11.6$ & $105.7\pm11.6$ & $92.0\pm9.7$ & $95.5\pm11.9$ & $87.9\pm10.5$ \\
    ID2 (Isophoto) & $-37.5\pm32.1$ & $9.0\pm11.5$ & $26.7\pm11.2$ & $2.6\pm11.9$ & $7.4\pm13.7$ & $51.9\pm15.0$ & $115.6\pm10.4$ & $75.5\pm13.7$ & $110.7\pm6.3$ & $88.2\pm8.3$ & $75.3\pm5.8$ & $86.9\pm9.3$ & $92.6\pm6.9$ \\
    ID2 (Kron) & $-23.4\pm20.0$ & $4.1\pm11.4$ & $10.4\pm7.6$ & $7.9\pm11.2$ & $23.3\pm13.3$ & $53.4\pm13.8$ & $137.0\pm12.4$ & $99.8\pm13.2$ & $98.4\pm6.9$ & $87.0\pm7.8$ & $73.7\pm7.2$ & $84.9\pm8.3$ & $64.9\pm6.9$ \\
    A.2 & - & - & - & - & - & $34.7\pm4.5$ & $73.8\pm3.4$ & $54.5\pm4.0$ & $47.4\pm3.5$ & $44.5\pm3.7$ & $42.5\pm3.5$ & $41.3\pm3.4$ & $40.2\pm3.7$ \\
    B.2 & - & - & - & - & - & - & $17.2\pm5.6$ & - & $19.2\pm5.9$ & $18.7\pm5.8$ & $16.3\pm6.0$ & - & $13.3\pm5.4$ \\ 
    C.2 & - & - & - & - & - & $23.1\pm8.5$ & $26.5\pm9.2$ & $47.0\pm8.3$ & $31.9\pm9.2$ & $17.1\pm8.1$ & $28.8\pm9.2$ & $27.0\pm9.4$ & $30.2\pm9.3$ \\ 
    \hline
    \end{tabular}
    \hspace{6cm}
    }
    \begin{tablenotes}  
    \footnotesize
    \item Notes. Observed fluxes, not corrected for magnification. For ID1 and ID2, we present fluxes measured with customized, total isophotal, and Kron apertures (Section \ref{subsec:photometry}). We use customized aperture (total isophotal) fluxes of ID1 and ID2 to estimate photometric redshift (stellar population). For individual clumps, the fluxes are derived with \texttt{GALFIT} fitting.
    \end{tablenotes}
    \label{tab:photometry}
    \end{center}
\end{table*}
\end{longrotatetable}

\end{document}